\documentclass[a4paper,oneside,12pt]{article}

\usepackage{graphicx}
\usepackage{pstricks}
\usepackage{pst-pdf}
\usepackage[margin=2.5cm]{geometry}
\usepackage{setspace}
\usepackage{cite}
\usepackage{subfigure}
\usepackage[font=small]{caption}
\usepackage{multicol}
\usepackage{pdflscape}
\usepackage{longtable}
\usepackage[parfill]{parskip}

\usepackage[normalem]{ulem}
\usepackage{booktabs}
\usepackage{multirow}
\usepackage{stfloats}
\usepackage{amssymb}
\fnbelowfloat 

\def\xmm{{\it XMM-Newton~\/}}

\def\Msun{\hbox{$\rm ~M_{\odot}$}}

\def\ergsec{$\rm erg~s^{-1}$}

\def\chisq{{\chi^{2}}}
\def\rchi{{\chi^{2}_{\nu}}}

\def\H0{{\rm ~km~s^{-1}~Mpc^{-1}}}

\def\la{\mathrel{\hbox{\rlap{\hbox{\lower4pt\hbox{$\sim$}}}{\raise2pt\hbox{$<$}}}}}
\def\ga{\mathrel{\hbox{\rlap{\hbox{\lower4pt\hbox{$\sim$}}}{\raise2pt\hbox{$>$}}}}}

\def\d25{D$_{25}$}

\def\hi{H {\small I}$~$}
\def\hii{H {\small II}$~$}
\def\oii{O {\small II}}
\def\oiii{O {\small III}}

\def\.25{0.25 keV\thinspace}

\def\liners{low-ionization nuclear emission-line regions\thinspace}
\def\llagn{low-luminosity active galactic nucleus\thinspace}
\def\llagns{low-luminosity active galactic nuclei\thinspace}
\def\agn{active galactic nucleus\thinspace}
\def\3xmm{3XMM-DR4\thinspace}

\def\a{$\times 10^{38}$}

\def\d{$\times 10^{41}$}
\def\e{$\times 10^{42}$}

\begin{document} 
\bibliographystyle{amsplain}
\singlespacing
\title{\textbf{Sleeping Giants?}} 
\date{\textbf{April 2014}} 
\author{\textbf{Ivayla Emilova Kalcheva} \vspace{1cm}\\  Master's thesis, Durham University, UK \\Programme of study: MPhys Physics and Astronomy\\ Supervisor: Dr T. P. Roberts}

\maketitle

\begin{abstract}          

New X-ray observations allow us to probe closer to the central engine of active galactic nuclei (AGN) than previously possible. This new information, combined with the existing census of nearby galaxies, enables us to study low-luminosity AGN (LLAGN), potentially accreting in a radiatively inefficient regime.
 In this project, the X-ray nuclear  properties of a  sample of  bright nearby  galaxies are explored. This is done by matching their comprehensive optical spectroscopic classification to the latest available \xmm catalogue - \3xmm. The good coverage ($\approx$ 38\%) ensures that a statistically representative sample is investigated in the project. All  nuclear and morphological subsets found within the original  sample of 486 galaxies are encompassed, but early-type galaxies and galaxies with optical AGN features are favoured.
The results from the investigation of the properties of our cross-matched sample, such as  source offsets from the nuclear positions, X-ray luminosities, black hole masses, Eddington ratios, are overall consistent with the presence of a large fraction of X-ray - detected \llagns.

The X-ray - detected galaxies within our \hii and transition-LINER subsets are of particular interest for this project, as they could harbour  LLAGN missed by optical spectroscopic selection. 
The properties of these nuclei are explored by X-ray spectral fitting of available \xmm observations.

Standard spectral models are employed to determine whether the nuclei within our selected sample are truly powered by starburst, or by an optically undetected LLAGN. The majority of the examined spectra are  found to be  consistent with a thermal plasma model and an underlying power-law continuum, and the relative contributions of a potential LLAGN and of thermal emission to the source luminosity are presented.  
In the case that these spectra are not highly contaminated, as confirmed for some of the nuclei by examining available higher-resolution \textit{Chandra} images, we classify $\approx$ 43\% of the \hii nuclei and 40\% of the transition-LINER nuclei as optically undetected LLAGN-candidates. 
Our classification is found to be reasonably successful, given the project limitations, when compared to X-ray and multi-wavelength results found in literature for many of the nuclei.

\end{abstract}

\newpage

\tableofcontents

\newpage

\section{Introduction and Theory}\label{intro}
\subsection{Motivation}

In the local Universe, most supermassive black holes (SMBHs) in galactic nuclei are 'sleeping' - they are dim and dormant in comparison to old and distant quasars. Low-luminosity active galactic nuclei (LLAGN) are the dominant mode of nuclear activity at low redshift \cite{ho:2008}. The X-ray band provides the clearest view of the central regions of these relatively timid giants. X-rays are emitted in vast quantities by the accreting SMBHs and the advent of satellite X-ray observatories significantly contributed to developing theoretical models to understand the physical mechanisms driving AGN and LLAGN. 

To explore the behaviour of weakly accreting SMBHs in the nearby parts of our Universe, a big sample of galaxies with X-ray detections is required. This project is concerned with analysing the X-ray properties of a sample of nearby galaxies, using data from \textit{XMM-Newton} (the latest, \3xmm catalogue) to investigate LLAGN activity, the specifics of the accretion process and the accretion rates. We base this work on a sample of 486 galaxies, the optical spectroscopic properties of which were explored by Ho, Sargent and Filippenko \cite{ho:1995}\cite{ho:1997b}, of which 38\% are found in the 3XMM-DR4 field of view. The project focuses on X-ray spectral analysis of nuclei optically classified as \hii and transition-LINERs, which have been less extensively studied in comparison to pure LINERs and Seyferts \cite{exploring:2010}.

\subsection{Active Galactic Nuclei - high-state accretion onto a SMBH}\label{variable}
The central compact regions of active galaxies have a higher than normal luminosity output, covering most or all wavebands, and the spectra are partly non-thermal, in this way differing from stellar emission. AGN are variable, as a compact object (less than one light-day across) emits a thousand times the energy of a normal galaxy. In the case of the most powerful and distant AGN - the quasars - the nuclei appear star-like and the host galaxy is almost completely outshone \cite{exploring:2010}. 
Some of the more exotic first explanations for the small but very bright region that was emitting the radiation included annihilation \cite{burbidge:1956}, white holes \cite{neeman:1965} and quark stars \cite{burbidge:1967}. There is a variety of AGN types, based on brightness and the classification schemes used \cite{ho:2008}.

The modern explanation for the underlying physical mechanism causing the observed emission is accretion onto a supermassive black hole ($M_{BH} > 10^{6}-10^{9}\Msun$), the most efficient known means of converting matter to energy ($\epsilon\sim 10\%$, up to $\sim 40\%$ for a spinning black hole) \cite{bell:1969}.    This is supported by observation, e.g. short-time variability, Doppler velocity broadening in UV/optical emission lines, collimated jets, etc. \cite{exploring:2010}. Moreover, it is widely accepted that most if not all galaxies have a central SMBH (in all bulges, sometimes even in bulgeless, dwarf galaxies), but it is still unclear whether the SMBH always undergoes an active stage \cite{ho:2008}. 

\begin{figure*}[!t]
\centering
\includegraphics[width=0.75\textwidth]{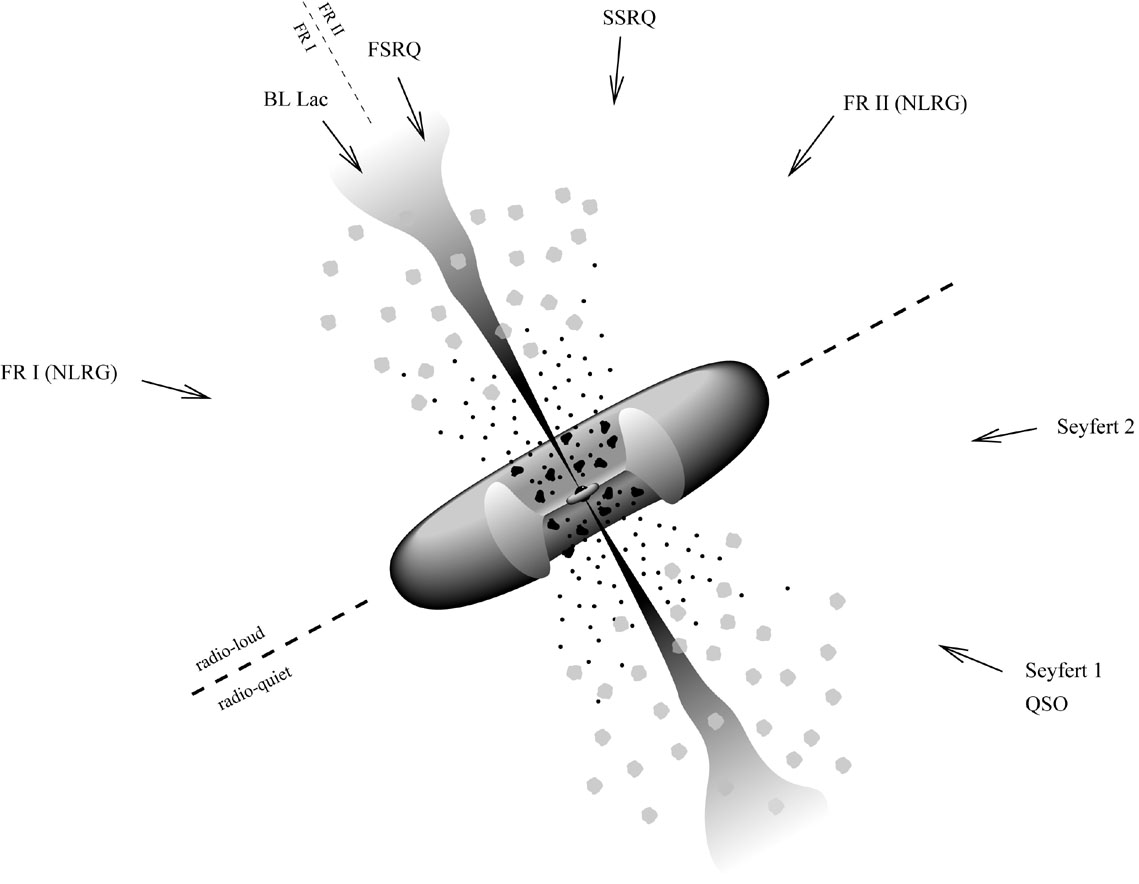}\label{fig:unified}
\caption{AGN unification \cite{torres:2004}.}
\label{fig:unification}
\end{figure*}

AGN unification models state that all active galactic nuclei share the same central engine and that the underlying mechanism is scale-invariant - AGN and quasars differ in some of their observable characteristics due to their orientation towards the observer's line of sight \cite{ho:2008} (Fig. \ref{fig:unification}). 
From fluorescence in the colder material in the surroundings of the nucleus, irradiated by X-ray emission from the source, a Compton reflection spectrum and a Fe K$\alpha$ line are commonly produced. The Fe K$\alpha$ line  can be the most prominent feature in the range 2-10 keV found in the spectra of AGN, with equivalent widths of $\gtrsim$ 1 keV \cite{ogle:2000}. If the line of sight intercepts the dusty torus characteristic for AGN, a large portion of the emission is absorbed. 

\subsection{Low-luminosity AGN - scaled-down AGN, or differences in mechanism?}
 
\begin{figure*}[!t]
\centering

\includegraphics[width=0.9\textwidth, height=8.5cm]{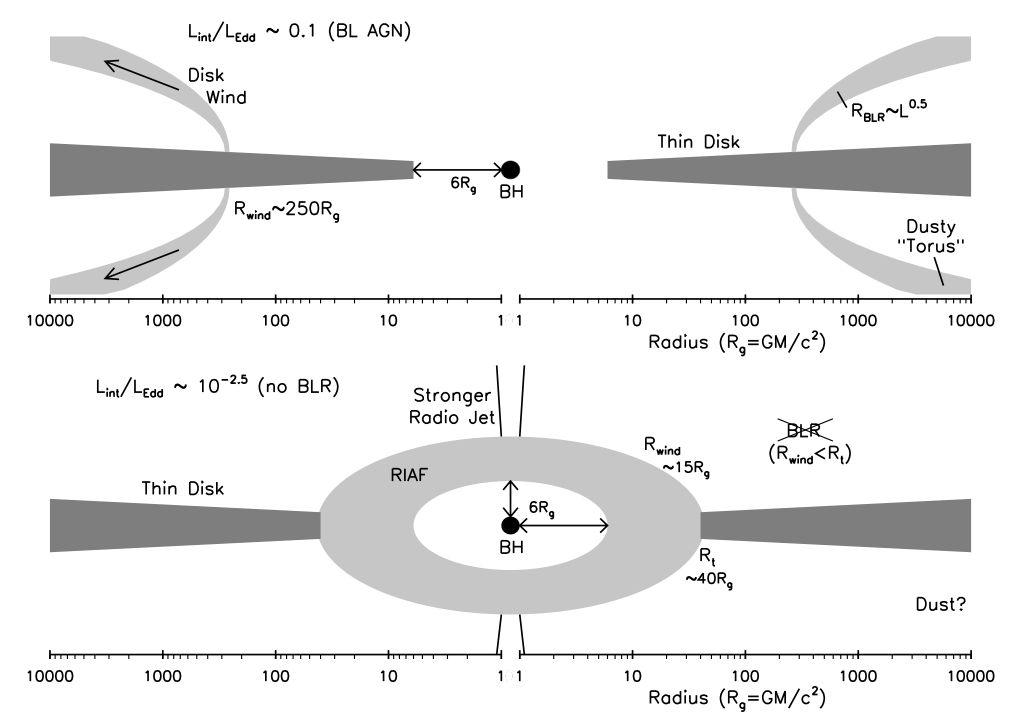}\label{fig:accretion}

\caption{Image illustrating the main differences between AGN and LLAGN with RIAF \cite{trump:2011}.}
\label{fig:riaf}
\end{figure*}

Observing activity at different redshifts has shown that nearby active galaxies each on average generate 1\% of the radiation from quasars in early times and distant AGN. The total output of these \llagns (LLAGN), however, is 10\% of the emission by quasars.  This indicates that there are many more moderate active black holes now than were quasars. The phenomenon is known as cosmic downsizing \cite{alexander:2007}, and is observed also for starburst (\hii) galaxies, where there is drop in star formation rates (SFR) over cosmic time.  LLAGN are found in $>$ 30\% of nearby galaxies \cite{exploring:2010}. Early models to explain the intrinsic dimness of LLAGN assumed that the central engine of LLAGN is an exact scaled-down version of AGN, but this hypothesis falls apart under the pressure of observational evidence \cite{ho:2008}. Apart from the much lower luminosity of LLAGN ($L_{Bol}/L_{Edd} < 10^{-2}$), their spectra in the optical and UV lack a 'big blue bump' - a feature characteristic for thermal emission from a Compton optically thick, geometrically thin accretion disc in an \agn, and have a 'red bump' instead (a mid-IR peak and a steep spectrum fall-off due to the small disc extent and the lower accretion efficiency). Other missing AGN features are the obscuring dusty torus and the broad line region (BLR). 
Therefore, LLAGN cannot be incorporated in the standard AGN unification model \cite{exploring:2010}. The current best physical model to explain these features consists of three components: a radiatively inefficient accretion flow (RIAF), a jet/outflow, and a truncated thin disc on the outside \cite{ho:2008} (Fig. \ref{fig:riaf}). A RIAF can be, for example, due to an advection-dominated accretion flow, or another dissipative process such as convection. This different mode of accretion results from low accretion rate, for example due to insufficient 'fuel'. Jet-only models exist - with jets instead of accretion \cite{exploring:2010, griffiths:1998}.

\subsection{The Ho, Filippenko and Sargent optical spectral classification of LLAGN}

\begin{figure*}[!ht]
\centering
\hbox{
\includegraphics[width=9cm]{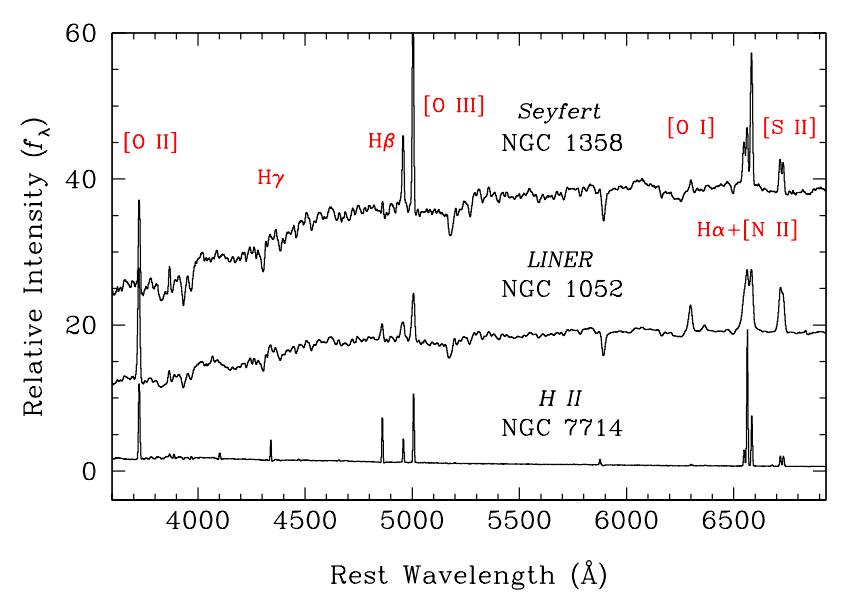}
\includegraphics[width=6.5cm]{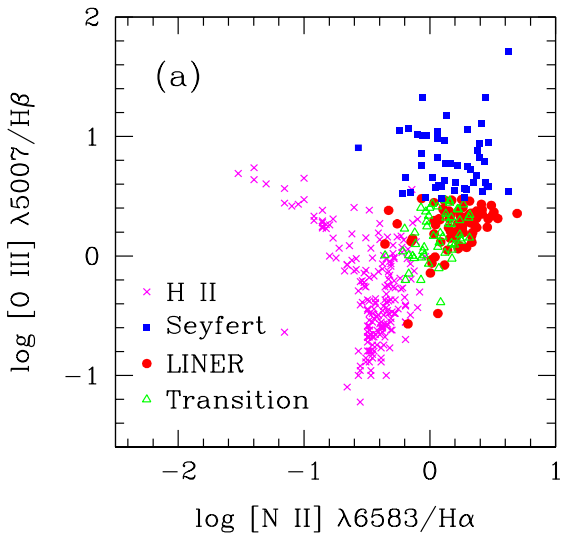}
}
\caption{Left: using optical spectroscopy to classify \llagns \cite{ho:1995}. Right: diagnostic diagram of LLAGN \cite{ho:1997a}.}
\label{fig:ho}
\end{figure*}

Spectral classification is used to put LLAGN types into order - the best-established method is to take the ratio of prominent emission lines, in order to determine whether the excitation source is an active nucleus or stellar photoionization. In this way, high-ionization (Seyfert) nuclei are those which have an optical ratio [\oiii]/[\oii] $>$ 1, and  \liners (LINERs) have [\oiii]/[\oii] $<$ 1 \cite{veilleux:1987} (Fig. \ref{fig:ho}). Both classes can be further divided into type 1 and type 2 nuclei, determined by the presence/lack of broad emission lines in the spectra. Diagnostic diagrams feature a region of ambiguous cases - dubbed as Transition nuclei, with composite LINER/\hii spectra  \cite{ho:1997a, ho:1997b, ho:2008} (Fig. \ref{fig:ho}). 
Optical spectroscopic surveys (Palomar, SDSS) give a lower boundary of the total number of nearby active galactic nuclei\cite{zhang:2009}. LLAGN could also be hiding in galaxies with starburst (\hii) regions\cite{jackson:2010}. 

The morphology of the host galaxy is important - activity in nearby early-type galaxies is often due to the presence of a Seyfert, LINER or transition nucleus, whereas nuclei in late-type galaxies seem to be dominated by star-forming regions. 
The detailed optical survey by Ho, Filippenko and Sargent \cite{ho:1997a, ho:1997b, ho:1997c} of the sample of 486 (mostly) northern galaxies comprises an excellent basis for a search for LLAGN in another waveband, as it is representative of the nearby bright galaxy population. Adopting their classification permits the division of the studied galaxy sample into optical nuclear and morphological subsets.

\subsection{Reasons for using X-ray data to look for LLAGN}\label{reasons}

The optical and infrared photons reaching our detectors have not arrived directly from the galactic nucleus - they are absorbed and re-emitted (even sometimes with a change in wavelength) by gas and dust clouds around the central region. Radio and high-energy waves, on the other hand, pass unruffled through these matter clouds, and studies in those bands are not as strongly affected by  contamination by star-forming regions \cite{ho:2008}.
While not all AGN are radio-loud, all emit X-rays, and even the \llagns emit them in vast quantities compared to other objects in the host galaxies. Another major advantage is that the contrast is much better in X-rays, compared to detected photons in other wavelengths, as there are much fewer X-ray sources \cite{ptak:2000}.

The processes responsible for the X-ray emission are both thermal and non-thermal. Thermal processes arise in optically thin plasma ($T > 10^{6}-10^{7}$K) - these include bremsstrahlung from the host galaxy, line emission, as well as direct emission from the accretion disc. 
Non-thermal  processes include synchrotron radiation and inverse Compton scattering, caused by the hot accretion disc and the jets \cite{exploring:2010}. Thus, X-rays provide one of the best ways to probe very close to the nucleus and allow its identification. 

Combining X-ray surveys with studies in other wavebands can not only improve the statistics, but also our knowledge of the accretion mechanism \cite{zhang:2009}. The brightest nearby objects (Seyferts, LINERs) have been quite extensively studied in the optical and, using the data accumulated since the first X-ray missions, in X-rays as well \cite{ho:2008, exploring:2010}.  For the ambiguous cases, such as optically-classified transition LINERs and \hii nuclei, however, the most up-to date X-ray data are needed - with resolution sufficient to reach as close to the central black hole as possible, thus minimising contamination by background sources and non-nuclear sources in the host galaxy.

\subsection{\textit{XMM-Newton} and the 3XMM-DR4 catalogue}
\subsubsection{The observatory}\label{xmm}

Such an opportunity is provided by the data in the latest \textit{XMM-Newton} catalogue, \3xmm. \textit{XMM-Newton}, launched in 1999 by the European Space Agency (ESA), is the the satellite which supplies the data for the catalogue. The observatory has a 30 arcmin field of view and carries four co-aligned telescopes: 3 high-throughput X-ray telescopes and an optical/UV monitor telescope. The science instruments on board include\footnote{http://xmm.esac.esa.int/external/xmm\_user\_support/documentation/uhb\_2.1/node1.html}:

- The European Photon-Imaging Camera (EPIC). It consists of one pn and two MOS (Metal Oxide Semi-conductor) CCD cameras, which allow timing studies, X-ray imaging, X-ray photometry and moderate-resolution spectrometry to be performed. Table \ref{epic} includes more details for the EPIC cameras. The relative pointing accuracy of each camera is $<$ 1.5" across the field of view. Between all cameras the relative astrometry is also $<$ 1"- 2", and the absolute pointing accuracy can be reduced to $\sim$ 1.5".

\begin{figure*}[!b]
\centering
\includegraphics[width=0.5\textwidth]{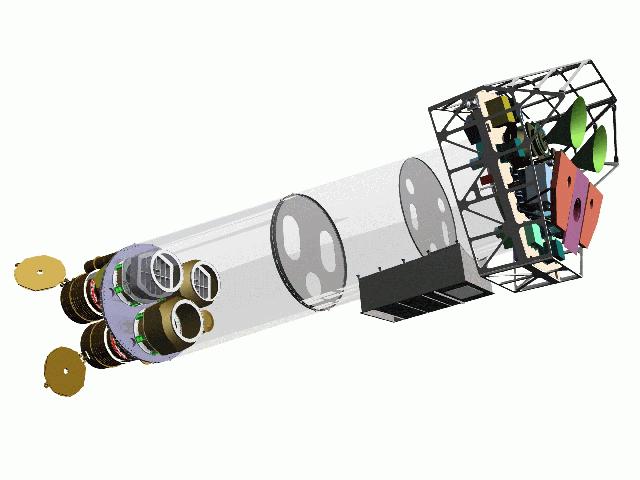}
\caption{The $\textit{XMM-Newton}$ payload: mirror modules (two of which with reflection gratings) situated at the lower left. Right: the focal X-ray instruments and the two RGS detectors. The figure is from the \textit{XMM-Newton} website.
}
\label{fig:xmm}
\end{figure*}

- Two Reflection Grating Spectrometers (RGS) for high resolution X-ray spectroscopy. They are co-aligned with MOS1 and MOS2, with all four instruments placed on the same telescope. The gratings disperse the wavelengths at different angles and the spectrum is collected and analysed by linear arrays of CCD chips similar to those in the MOS cameras.

- Optical Monitor (OM) - the first on an X-ray observatory. The OM helps achieving a good alignment of the sources in the optical and X-ray wavebands\footnote{http://sci.esa.int/xmm-newton/47370-fact-sheet/}. Figure \ref{fig:xmm} shows the \textit{XMM-Newton} payload.

\begin{table*}[!t]
\centering
\begin{tabular}{|lcc|}
\hline
\textbf{Instrument}	& \textbf{pn}&\textbf{MOS}\\ 
\hline
\hline
Pixel size & 150 $\mu$m & 40 $\mu$m \\
PSF (FWHM)& 6"&5" \\
Field of view	& 30'  & 30' \\
Bandpass  &0.15-15 keV &0.15-12 keV \\
Temporal resolution & 0.03 ms & 1.5 ms\\
Spectral resolution (1 keV) & $\sim$ 80 eV & $\sim$ 70 eV \\
Spectral resolution (6.4 keV) & $\sim$ 150 eV & $\sim$ 150 eV \\
\hline
\end{tabular}
\caption{Some of the characteristics of the EPIC cameras.}
\label{epic}
\end{table*}

The observatory is capable of simultaneous (and independent) operation of all instruments, as well as simultaneous optical/UV with X-ray observations. The latter allows X-ray sources to be monitored and identified against an optical image of the surrounding field. Moreover,
\textit{XMM-Newton} has good angular resolution (Table \ref{fig:xmm}) and is highly sensitive (due to having the largest effective area of a focusing X-ray telescope built so far), with special notice given around 7 keV. Photometry allows variability of X-ray sources over time to be detected. Spectral analysis is performed from the very soft to the hard X-ray band ($\sim$ 0.1 - 15 keV). The high sensitivity of the instruments makes the satellite very well suited for spectral analysis. The excellent field of view and high throughput spectroscopy means that \textit{XMM-Newton} is ideal for survey studies of the bright X-ray sources in nearby host galaxies.  

Whenever the angular resolution of \textit{XMM-Newton} is insufficient to resolve individual point sources, the \textit{Chandra} X-ray telescope offers higher angular resolution of $\sim$ 0.2" (PSF at FWHM)\footnote{http://xmm.esac.esa.int/external/xmm\_user\_support/documentation/uhb\_2.1/node86.html\#2849}. 

\subsubsection{The catalogue}

The 3XMM-DR4 X-ray source catalogue 3, released on 23 July 2013, is the latest and best (so far) \textit{XMM-Newton} catalogue. It contains 50\% more detections than its predecessor, covering 5 X-ray energy bands with a total range of 0.2 - 12.0 keV. The net sky area covered independently is 794 square degrees, with 531 261 X-ray detections in 372 728 unique sources.  Variability of sources during a single exposure is determined by photometry 
- after a $\chisq$ variability test to the time series, 4612 detections have been flagged as variable.  This is very useful, as short-time variability is a tell-tale for the presence of a compact source. The pn, MOS1 and MOS2 fluxes generally agree to $\sim$ 10\%. 

The data are pipeline-processed with the \textit{XMM-Newton} Science Analysis Software\footnote{http://xmmssc-www.star.le.ac.uk/Catalogue/3XMM-DR4/UserGuide\_xmmcat.html\#Catalogue}. The catalogue is wider than data from any previous survey in the hard band (2-12 keV) and enables good multi-wavelength source matching, giving a good resource for our search of LLAGN.  

\subsubsection{X-ray spectral data}\label{sfr}

\begin{figure*}[!b]
\centering
\includegraphics[width=0.5\textwidth]{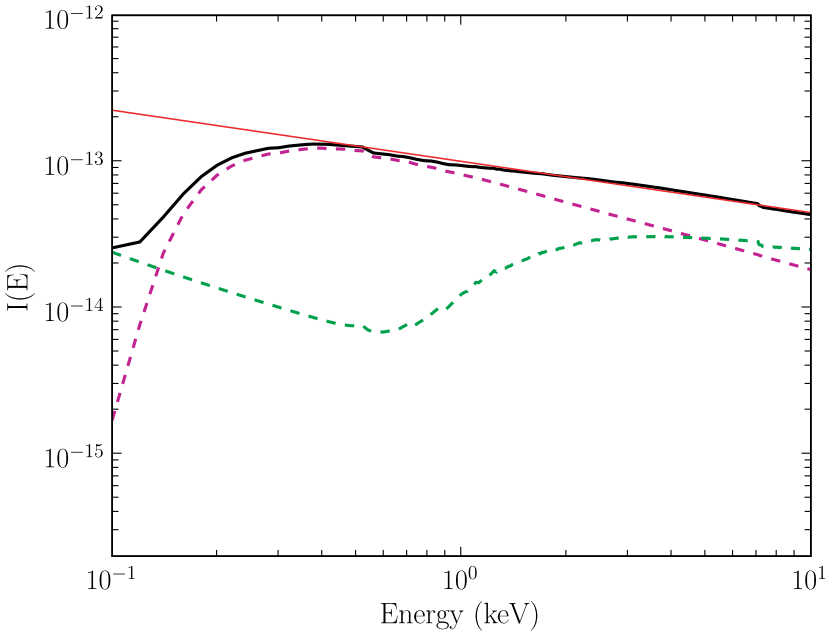}
\caption{Average \textit{XMM-Newton} spectrum of X-ray fits to a uniform sample of 102 nearby AGN. Red line - power-law component; dashed lines - simple and complex absorption components. Parameters: z = 0.03, 
energy range 0.1-12 keV , $\Gamma$ = 1.369 $\pm$ 0.004 \cite{winter:2009}.}
\label{fig:avg}
\end{figure*}

The first step of  X-ray spectral analysis is to identify the source and specify a source and background region, from which the observed source and background spectra are extracted. The preliminary spectra contain the raw number of counts per instrument channel. This data are then converted to flux per unit energy using detector-specific response files. Typically, the input spectra are weighted by telescope area and detector efficiency versus energy -  they are a convolution of the response matrix function (associating the photon energy to each instrument channel) and the ancillary response function (the quantum efficiency of the detector) - RMF $\otimes$ ARF \cite{gandhi:2005}. This folding of the model spectrum through the instrument response is more useful than attempting to deconvolve the response and obtain a 'true' source spectrum, as the response files give the \textit{probability} of a photon of a specific energy to be associated to a specific channel. Therefore, folding the model spectra through the response and fitting to the data is the standard method used to analyse X-ray spectra, via software such as {\sc{xspec}}\footnote{http://heasarc.nasa.gov/xanadu/xspec/} \cite{gandhi:2005, ho:2008}. 

\begin{figure*}[!ht]
    \centering
 \includegraphics[width=0.45\textwidth]{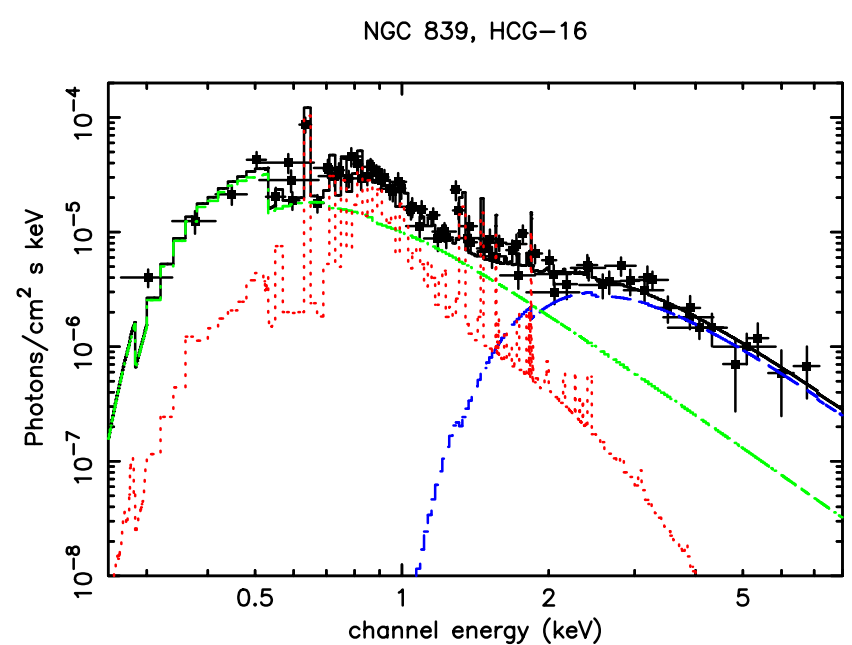}\label{fig:transition}
\includegraphics[width=0.45\textwidth]{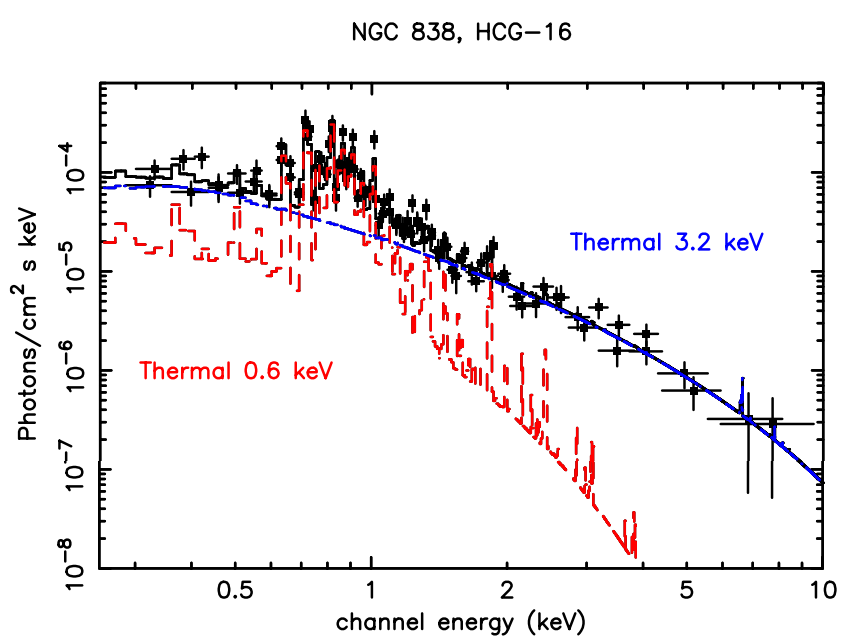}\label{fig:starburst}
\caption{Comparison between example EPIC-MOS X-ray spectra: a transition LINER (NGC 839), showing both starburst emission and emission from an obscured LLAGN, and a \hii nucleus (NGC 838), where only emission from the starburst is present and hard X-ray emission may be from unresolved X-ray binaries \cite{turner:2001}.}
\label{fig:vs}
\end{figure*}

The X-ray properties of LLAGN can be investigated in more detail by performing spectral fitting to obtain model parameters such as X-ray luminosity, photon index ($\Gamma$), absorption (\hi column density) of the host galaxy, plasma temperature ($k_{B}T$).  In the case of AGN and LLAGN, the X-ray spectra are continuous, generally power-law dominated (Fig. \ref{fig:avg}), with a photon index typically in the range $\sim$ 1.5 - 2.5 \cite{ishibashi:2010}. 
 
The exhibited soft (photoelectric) absorption is due to the Milky Way ISM and the material in the host galaxy. This is known as a 'cold absorber', suppressing the main power law continuum to a certain extent, depending on the column density in the line of sight. In the case of Compton-thick (type 2) AGN, the column density can exceed 10$^{24}$ cm$^{-2}$ \cite{maiolino:2007}. 

An important test to determine whether the observed X-ray central source has AGN or starburst origin is to differentiate between point-like and spatially extended emission. 
An extended X-ray emission region would suggest that all or part of the X-ray radiation is produced in a  star-forming region. Moreover, low-redshift pure starburst galaxies generally do not exhibit unresolved X-ray nuclei in the 2 - 10 keV range \cite{ptak:2000}. However, this test is not always definitive, as a compact source does not rule out the presence of a compact nuclear starburst region on scales smaller than a few kpc, when the ambiguity on the source distance (the main source of uncertainty) is taken into account \cite{ptak:2000}. Late type galaxies (rich of gas and young massive stars) in particular are affected by this, due to populations of high-mass X-ray binaries (HMXBs) found in such compact star-forming regions \cite{winter:2009}. 
Spectral data for the \hii and Transition nuclei in our sample was available from the \textit{XMM-Newton} Science Archive\footnote{http://nxsa.esac.esa.int/nxsa-web/\#search}. Typical EPIC-MOS transition-LINER and starburst X-ray spectra are shown in Fig. \ref{fig:vs}. The transition-LINER exhibits both starburst emission and emission from an obscured LLAGN, whereas only starburst emission is present in the \hii spectrum, and its hard X-ray emission could be due to unresolved X-ray binaries.

\section{Methods}\label{method}
\subsection{Selecting our X-ray sample by matching to the Ho, Filippenko  and Sargent optical 
classification}

As explained in Section \ref{intro}, a statistically significant, representative sample of nuclear X-ray sources was needed to study the activity in nearby galactic centres. 
The X-ray sample for this study was obtained by cross-matching the galaxies from the Ho, Filippenko and Sargent optical sample with the \3xmm catalogue. This was performed using the interactive graphical viewer and editor for tables {\sc topcat}\footnote{http://andromeda.star.bris.ac.uk/$\sim$mbt/topcat/}. As the nuclear positions quoted in the Ho optical sample carry errors up to 7.5" \cite{ho:1995}, we took the galaxy coordinates from the Nasa Extragalactic Database (NED\footnote{http://ned.ipac.caltech.edu/}). 

To cross-correlate the two datasets, all X-ray sources separated by less than an estimated threshold radius from the optical centre were counted as matches. It was possible in {\sc topcat} to visually inspect randomly chosen  Sloan Digital Sky Survey (SDSS\footnote{http://www.sdss.org/}) images of the host galaxies as a check to confirm that the matched sources are found at the optical centre of the host galaxy.

\begin{figure*}[!t]
    \centering
    \hbox{
\includegraphics[width=0.426\textwidth]{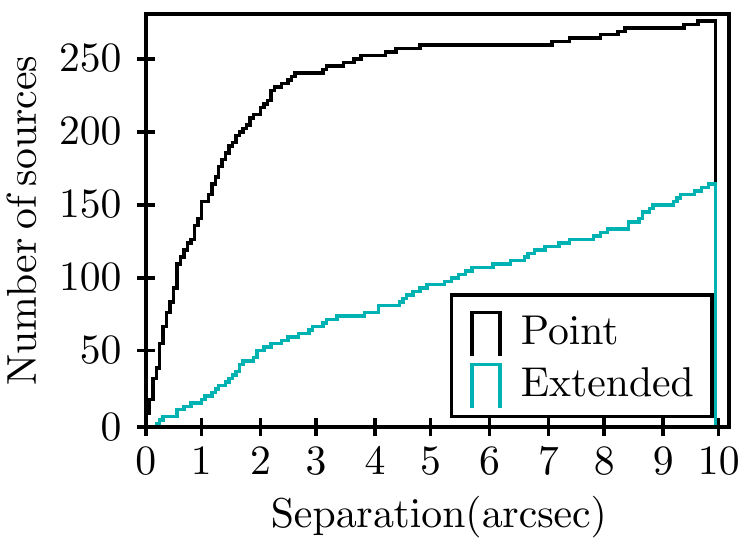}\label{fig:cum}
\includegraphics[width=0.581\textwidth]{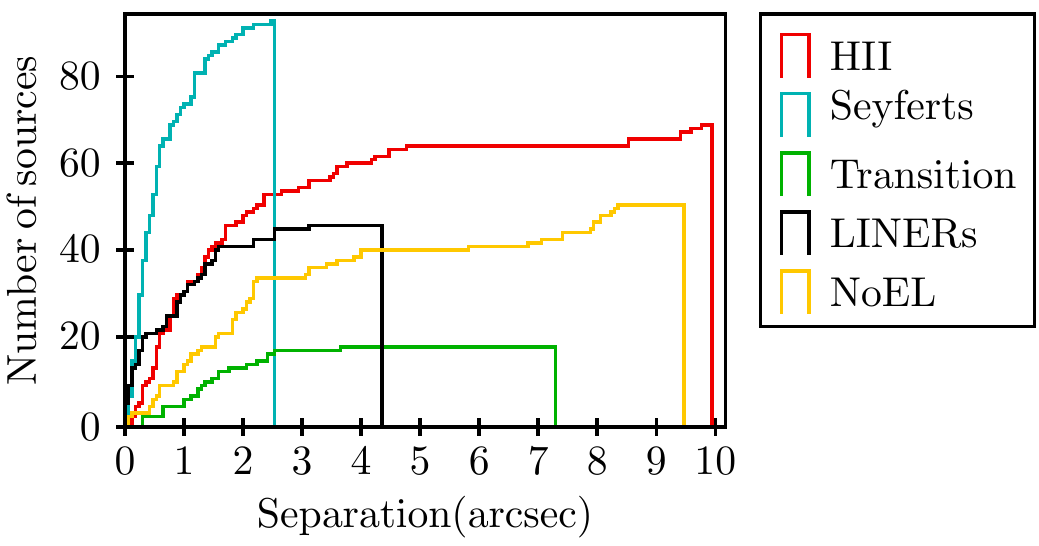}\label{fig:sub}
}\caption{Source number versus separation cumulative plots to estimate the appropriate threshold match radius for the X-ray sample. Left: showing the difference between X-ray point and extended sources. Right: point sources divided into nuclear subsets using the optical classification scheme.}
\label{fig:separation}
\end{figure*}

Figure \ref{fig:separation} shows the cumulative plots used to calculate the fraction of X-ray sources  at different distances from the optical centre. The aim was to include all potential active SMBHs in the sample and to find the distance at which the 'background' starts. For these nearby galaxies, the 'background' - for a search radius up to 10"- 25" - would be in most cases made up of non-nuclear sources within the host galaxy rather than outside its optical extent and into the cosmic background. Thus, we estimate an appropriate match radius of $\sim$ 5 

arcseconds. This estimate is conservative given the position accuracy of \3xmm, to allow for the fact that optical centre positions are still not well quantified in all cases, i.e. due to the variety of galaxy shapes and orientations. The match radius should exclude any significant contamination by X-ray sources not belonging to the galaxy of interest - e.g. foreground Milky Way stars or background AGN. Inevitably, X-ray emitters of comparable luminosity to the LLAGN, residing in the dense central region or in star clusters nearby, still remain in the newly-formed X-ray sample. The distinction is especially difficult in the cases of weaker X-ray sources, e.g. 
 transition-LINER and \hii galaxies \cite{parejko:2008}.

Figure \ref{fig:separation} shows the cumulative plots for all point and extended sources in a 10" match sample.  The point source plots (presented for the whole sample in the left panel and separately for the nuclear subsets on the right) arise from a distribution with a strong peak of point sources very close to the optical nucleus, which would be expected for LLAGN. The plot for extended sources (left panel) does not exhibit such a peak, thus these sources appear unrelated to \llagns. Therefore, X-ray sources flagged as extended in \3xmm were kept for statistical purposes only and excluded from the search of LLAGN.

\subsection{Examining the cross-matched sample properties}

\subsubsection{Checking the coverage - what is the morphology and nuclear variety of our X-ray sample?}

The optical nuclear and morphological classification enables us to split our final matched X-ray sample into subsets, and in this way to check the coverage and, therefore, how representative the X-ray sample is. 

\begin{table*}[!ht]
\centering
\begin{tabular}{|lccc|}
\hline
\textbf{Morphology}	& \textbf{Optical sample}&\textbf{In field of view}& \textbf{Coverage} \\ 
\hline
\hline
E	& 57	    & 36	& 0.63	 \\
S0	& 88		& 28	& 0.32		 \\
S0/a-Sab	& 77		& 31	& 0.40		\\
Sb-Sbc	& 103		& 40	&0.39	\\
Sd-Scd	& 109		& 31	& 0.28		 \\
Sd-Sdm	& 19		& 4	& 0.21		\\
Sm-Im	& 21		& 10	& 0.48		 \\
I0	& 5		    & 4	& 0.80	 \\
Pec+S pec	& 7		    & 2	& 0.29	 \\
\textbf{All}	& \textbf{486}	    & \textbf{186}	&\textbf{ 0.38	} \\
\hline
\end{tabular}
\caption{Morphology and coverage of the galaxies in the optical sample with \textit{XMM-Newton}.}
\label{table:morph}

\centering 
\begin{tabular}{|l*{7}{c}|} 
\hline
\textbf{Nucleus type} & \textbf{Optical} &\multicolumn{3}{c}{\textbf{In field of view}} & \multicolumn{3}{c|}{\textbf{Coverage}} \\ 
 &  & \textbf{All}& \textbf{Detected} &\textbf{Point} & \textbf{All}& \textbf{Detected} & \textbf{Point}  \\ 
 \hline
 \hline
Seyferts	& 52	    &50 & 49&37	&0.96 & 0.94 & 0.71	 \\
pure LINERs	& 94		&40 & 40&29	&0.43 & 0.43 & 0.31  \\
transition LINERs & 65	&21 & 19&13	&0.32 & 0.29 & 0.20  \\
\hii	& 206		    &55 & 43&31	&0.27 & 0.21 & 0.15  \\
NoEL	& 69		    &20 & 18&12	&0.29 & 0.26 & 0.17  \\
\textbf{All	}& \textbf{486}		& \textbf{186}& \textbf{169}&	\textbf{122}&  \textbf{0.38} &\textbf{0.35}&	\textbf{0.25} 	\\
\hline 
\end{tabular}
\caption{Nuclear classification and coverage of the galaxies in the optical sample with \textit{XMM-Newton}. Not all galaxies in the satellite field of view had X-ray detections. In some of the detected galaxies, only extended X-ray sources were found
.}
\label{table:coverage}
\end{table*}

Knowing the field of view of \textit{XMM-Newton} (30 $\times$ 30 arcmin square patch seen at any given observation, Table \ref{epic}) and using the NED coordinates for the original sample of 486 galaxies, a 15-arcmin radius search from each centre of the \3xmm sources was performed to look for missing galaxies -  within the XMM field of view but without detected X-ray emitters. In total, out of the 186 optically-classified galaxies within the XMM field of view, no X-ray sources were found in the central regions of 17 galaxies. As some of these 'missing' galaxies were included as X-ray emitters with a central source in previous studies \cite{roberts:2000}, a second check indicated that some of them have X-ray source detections outside a 10" (and less than 25") search radius. Only 5 of those galaxies had a point source detection in the range 10" - 25".  

Therefore, no X-ray cores should be missing from our remaining sample. 

Table \ref{table:coverage} demonstrates that the coverage is sufficient to leave us with a representative of the nearby galaxy population sample, although the coverage of the early-type galaxies is better than for late-types. Also, the brightest LLAGN - Seyferts and pure LINERs, were favoured (e.g. while 96\% of the Seyferts are covered, we have only 27\% of the \hii nuclei). 

Some galaxies may be missing from these statistics - if unfortunate enough to have been in the lost edges between the circular search radius and the square field of view. (Although vignetting\footnote{http://xmm.esac.esa.int/external/xmm\_user\_support/documentation/uhb\_2.1/node21.html} would have affected the image edges anyway). 

\subsubsection{Calculating the  X-ray and bolometric luminosity}
Using the X-ray fluxes ($F_{X}$) from \3xmm (band 8, 0.2 - 12 keV) and the distances ($D_{L}$) to the galaxies in the X-ray sample, it was straightforward to estimate the X-ray luminosity ($L_{X}$):
\begin{equation}
L_{X}=4\pi {D}_{L}^{2}F_{X}~.
\label{eq:Lx}
\end{equation}

LLAGN have a tendency to be X-ray-loud. Therefore, a bolometric correction of 15.8 was required to obtain their bolometric luminosity $L_{Bol}$ - the total radiant energy in all wavelengths in all directions - instead of the conventionally used correction of $\approx$ 35 for luminous sources \cite{ho:2009a}:
\begin{equation}
L_{Bol}=15.8\times L_{X}~.
\label{eq:Lbol}
\end{equation}
The flux upper limits for the galaxies within the field of view without X-ray sources were obtained from FLIX\footnote{http://xmm.esac.esa.int/xsa/\#tools}, to compare detections and non-detections.

\subsection{Estimating the Eddington luminosity, black hole masses and accretion rates} \label{mass}
Using data for galaxy velocity dispersions \cite{hobh:2000}, we used the empirical relation (based on the $M-\sigma$ relation, \cite{magorrian:1998}):
\begin{equation}
{M}_{BH}=1.2\times{10}^{8}\Msun{\left(\frac{{\sigma }_{e}}{200\rm~km~s^{-1}}\right)}^{3.75}~,
\label{eq:Mbh}
\end{equation}
to estimate the central black hole mass, $M_{BH}$, of each galaxy in our sample. In the relation, $\sigma_{e}$ is the line-of-sight aperture dispersion inside the half-light radius, $R_{e}$ \cite{hobh:2000}.
The expected luminosity for accretion at the Eddington limit (the maximum luminosity possible when there is balance between the force of radiation and the gravitational force \cite{exploring:2010}), is obtained from 
\begin{equation}
{L}_{Edd}=\frac{4\pi G{M}_{BH}{m}_{P}c}{{\sigma }_{T}}~,
\label{eq:LEdd}
\end{equation}
 where $m_{P}$ is the proton mass, ${\sigma}_{T}$ is Thomson cross-section, $c$ is the speed of light, $G$ is the gravitational constant. For AGN, where $L_{X}$ is dominated by accretion, the luminosity is also proportional to the accretion rate. The mass accretion rate is calculated from
\begin{equation}
{\stackrel{.}{M}}_{Edd}=\frac{4\pi G{M}_{BH}{m}_{P}}{\epsilon c{\sigma }_{T}}~,
\label{eq:Medd}
\end{equation}
where $\epsilon$ $\sim$ 0.1 is the emissivity for a physically thin, optically thick disc \cite{exploring:2010}, used as an approximation.
 The Eddington ratio is defined as 
\begin{equation}
{\lambda }_{Edd}=\frac{{L}_{Bol}}{{L}_{Edd}}\propto \frac{{L}_{Bol}}{{M}_{BH}}~.
\label{eq:Eddratio}
\end{equation}
  
\subsection{Processing X-ray spectral data for the ambiguous cases - looking for LLAGN in Transition and \hii optically-classified nuclei}
 
\subsubsection{Fitting the \textit{XMM-Newton} spectra with {\sc xspec}} \label{fitting}

The X-ray properties of the optically classified \hii and transition-LINER nuclei were investigated in more detail by performing spectral fitting with the 
 {\sc xspec v12.8.1} software package\footnote{http://heasarc.nasa.gov/xanadu/xspec/}. 
Spectral datasets for the central source (within 5"  of the NED coordinates of the optical nucleus) were modelled. The pn, MOS1 and MOS2 spectra from each observation were fitted simultaneously, and a free normalization constant was added to account for the differences in flux calibration between the instruments.  The constant was set to 1 for MOS1 and left free in the  other two models. It did not vary significantly ($\sim \pm 10 \%$) between the three instruments. 

Only datasets with spectra for the central source with sufficient counts for a meaningful fit were used.  A minimum of 20 bins $\times$ 20 counts in total from the three instruments of $\xmm$ was a necessary lower limit for good  $\chisq$ statistics to be performed with {\sc xspec}. A minimum of $\sim$ 20 counts per bin ensured that the probability distribution of each data point was roughly Gaussian, with a $\sigma$ width matching the error bar. Response files\footnote{http://xmm.esac.esa.int/}  for each  of the EPIC detectors were used for the fitting. To ensure good data quality, the spectral data in the energy range of 0.3 - 10 keV were used, and bad channels were ignored. 

The main goal of the fitting was to distinguish between LLAGN candidates and non-AGN within the chosen optical subsets. Therefore, instead of searching for a more detailed fit for each central source, comparison between two main models was made -  absorbed power-law continuum (\texttt{const$\ast$tbabs$\ast$ztbabs$\ast$po}) and absorbed power-law continuum with thermal plasma (\texttt{const$\ast$tbabs$\ast$ztbabs$\ast$(apec+po)}). The additive component represents the source, whereas the modifying components model how radiation is affected on its way from the X-ray source to the detectors\footnote{https://heasarc.gsfc.nasa.gov/xanadu/xspec/XspecManual.pdf}. 

The \texttt{po} model is a simple photon power law and useful approximation of X-ray AGN spectra. It has the general form
$P\left(E\right)=K{E}^{-{\Gamma }_{eff}}$, where $K$ is the normalization constant (photons $\rm~keV^{-1}cm^{-2}s^{-1}$ at 1 keV), $E$ is the photon energy (keV), and $\Gamma_{eff}$ is the dimensionless photon index of the power law. 
$\Gamma_{eff}$ is not the true slope unless there is no intrinsic absorption.

The \texttt{apec} (Astrophysical Plasma Emission Code) model fits  a velocity- and thermally-broadened emission spectrum from collisionally-ionized diffuse gas. This was done as the quality of the \textit{XMM-Newton} spectra does not allow a clear distinction between collisionally and photo-ionized gas, and the collisionally ionized \texttt{apec} model is simpler. The \texttt{apec} model input parameters were: redshift, normalization, plasma temperature, and (fixed for all fits) metal abundances. 

The Galactic (foreground) absorption\footnote{http://heasarc.nasa.gov/cgi-bin/Tools/w3nh/w3nh.pl} in \texttt{tbabs} was fixed for the line-of-sight in each fit \cite{dickey:1990}, and \texttt{ztbabs} was used to obtain an estimate for the absorption from the environment of each host galaxy, taking into account the galaxy redshift. 
In all fits, ISM abundances were set using the \texttt{abund wilm} command.

The total absorbed flux was calculated with the \texttt{cflux} command applied to the whole model. The total intrinsic (unabsorbed) flux of the central source was obtained by applying \texttt{cflux} to the power law and thermal model, and the flux of the power law component was separately calculated to give an estimate for the contribution from a potential LLAGN. 

An improvement between models was considered statistically significant if $\chisq$ changed by 8 - 10 per degree of freedom. The model fits were considered reliable when the following criteria were fulfilled for the $\chisq$ statistics : 0.6 $ < \rchi <$ 1.5 and 
P($\rchi,\nu$) $>$ 5\%. The reliable fits were also expected to have physical parameters within established for the studied objects physically meaningful values - a temperature range 0 $ < kT <$ 2 keV, and a photon index range 0 $< \Gamma <$ 3 \cite{martin:2011}. Values close to the upper limit of  $\Gamma$ are found  for starburst and some LINER galaxies \cite{martin:2011, martin:2009}.

\subsubsection{Separation of the sample into LLAGN candidates and non-LLAGN}\label{checks}
A study of  \xmm and \textit{Chandra} data for a  representative sample of 82 nearby LINER galaxies \cite{martin:2009} was used for the initial separation of our sample into LLAGN and non-LLAGN candidates. Similar fitting techniques  - using an absorbed power law and absorbed power law with a \texttt{mekal} thermal component - were employed in the study. The AGN candidates in the LINER sample have an average threshold value  
 of $>$ 82\% power-law contribution to the 2-10 keV intrinsic luminosity. 
The average value is consistent for the \xmm and \textit{Chandra} fits: $\sim$ 83\% for the former and $\sim$ 81\% for the latter.  
Therefore, we assumed that below this value, the spectrum characteristics of our nuclear sources are more likely attributed to starburst emission rather than a LLAGN.

This separation was preliminary, as a contribution to the luminosity by the power law component could be an indicator for the presence not only of a potential \llagn, but also an X-ray binary or a source mixture.  
 The variability timescales of the sources can aid the classification: an X-ray binary would be varying on very short timescales (ms), thus noticeable during a single observation; a LLAGN varies for days or months; a combination of sources would lead to no observable variability \cite{ho:2008}. Simultaneous fitting with the same model was performed for the few central sources with available multiple observations to check for source variations on a timescale of days.  

Another test was the addition of a narrow \texttt{gauss} component to the models. This was done to constrain the possible presence of a  Fe K$\alpha$ line at 6.4 keV. The 99\% upper limit of its equivalent width was calculated to determine the potential contribution to the spectra.  No lines at lower energies were searched for, as they are expected to be hidden in the thermal component. 

When available, \textit{Chandra}-ACIS hard-band images (2-7 keV) were used for higher-resolution inspection of our LLAGN candidates. This was done to verify the presence of a nuclear point source and to check for cases of significantly contaminated nuclear sources in \textit{XMM-Newton} data.
Some of these images were already examined in studies \cite{zhang:2009, liu:2011}.  The remaining sources available from \textit{Chandra} were visually inspected with {\sc{cscview v.1.1.4}}\footnote{http://cda.cfa.harvard.edu/cscview/}. The files for the PSF at the source position were also available. The  extent of the nuclear source in the image was compared to the PSF (using {\sc{maxim dl v.5}}\footnote{http://www.cyanogen.com/maxim\_main.php}) to distinguish between point and extended emission: FWHM(Nucleus)-FWHM(PSF). 
In this way, it was possible to separate the sources in four categories (as suggested in \cite{zhang:2009, ho:2008}) : I -  dominant point source at the nuclear position; II - point source and extended emission; III - extended source only; IV - no source at nuclear position. 
The full table with the source classification is shown in the Appendix.

\section{Results}\label{results}

\subsection{Results for the galaxies with nuclear point X-ray sources in our cross-matched sample}

\subsubsection{Point source statistics}

Out of the 186 galaxies from the parent sample covered by \3xmm (and the 122 galaxies with point source detections, Table \ref{table:coverage}), 114 galaxies are found to contain nuclear point sources (i.e. within our 5" match radius), consistent with accreting black holes. Further 41 galaxies have only extended X-ray sources in the 5" nuclear region, and are therefore not considered as potential LLAGN candidates. 

All original optical nuclear subsets are represented in our remaining sample: nuclear point sources were detected in 76\% of the observed with \xmm Seyferts (38 out of 50),  $\approx$ 72\% of the pure LINERs (29 out of 40), $\approx$ 57\% of the transition-LINERs (12 out of 21), $\approx$ 47\% of the \hii nuclei (26 out of 55) and 45\% of the NoELs (9 out of 20).

As well as this, all morphological subsets defined in Table \ref{table:morph} are represented in our nuclear sample, but elliptical and early-type spiral galaxies are favoured.  Approximately 64\% (86 out of 135) of the early-type galaxies (E to Sb) have a point nuclear source, compared to 55\% (28 out of 51) for the later types (Sc to Sm, Irr). However, the presence of an X-ray point source at the optical nucleus is not unambiguous proof for a \llagn, especially for late-type spiral galaxies (see Section \ref{sfr}).

\subsubsection{Average X-ray luminosity}

\begin{table*}[!hb]
\centering
\begin{tabular}{|lcc|}
\hline
\textbf{Nuclear subset}	& \textbf{Count}& \textbf{Mean log }${L_{X}}$(\ergsec) \\
\hline
\hline
Seyferts	& 38	    & 41.36	$\pm$ 0.13	 \\
pure LINERs	& 29		& 40.26	$\pm$ 0.13		 \\
transition LINERs	& 12		& 39.80 $\pm$ 0.21		\\
\hii	& 26		& 39.23	 $\pm$ 0.10	\\
NoEL	& 9		& 37.72	$\pm$ 0.18		 \\
\hline
\hline
\textbf{All}& \textbf{114}	& \textbf{39.99	}$\pm$ \textbf{0.10}		\\
\hline
\end{tabular}
\caption{Average luminosities table.}
\label{table:lum}
\end{table*}

\begin{figure*}[!ht]
    \centering

\includegraphics[width=0.4\textwidth]{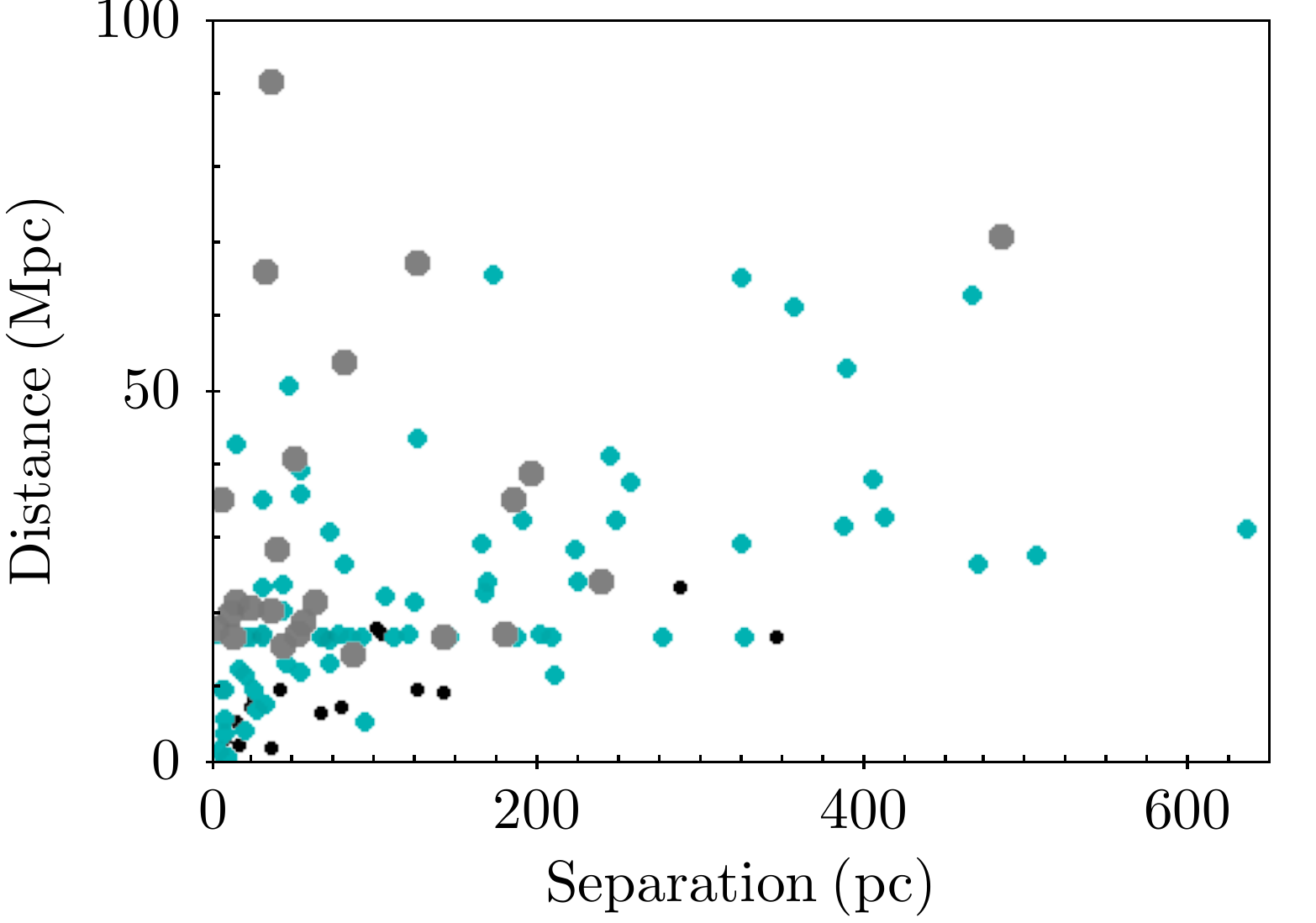}
\includegraphics[width=0.4\textwidth]{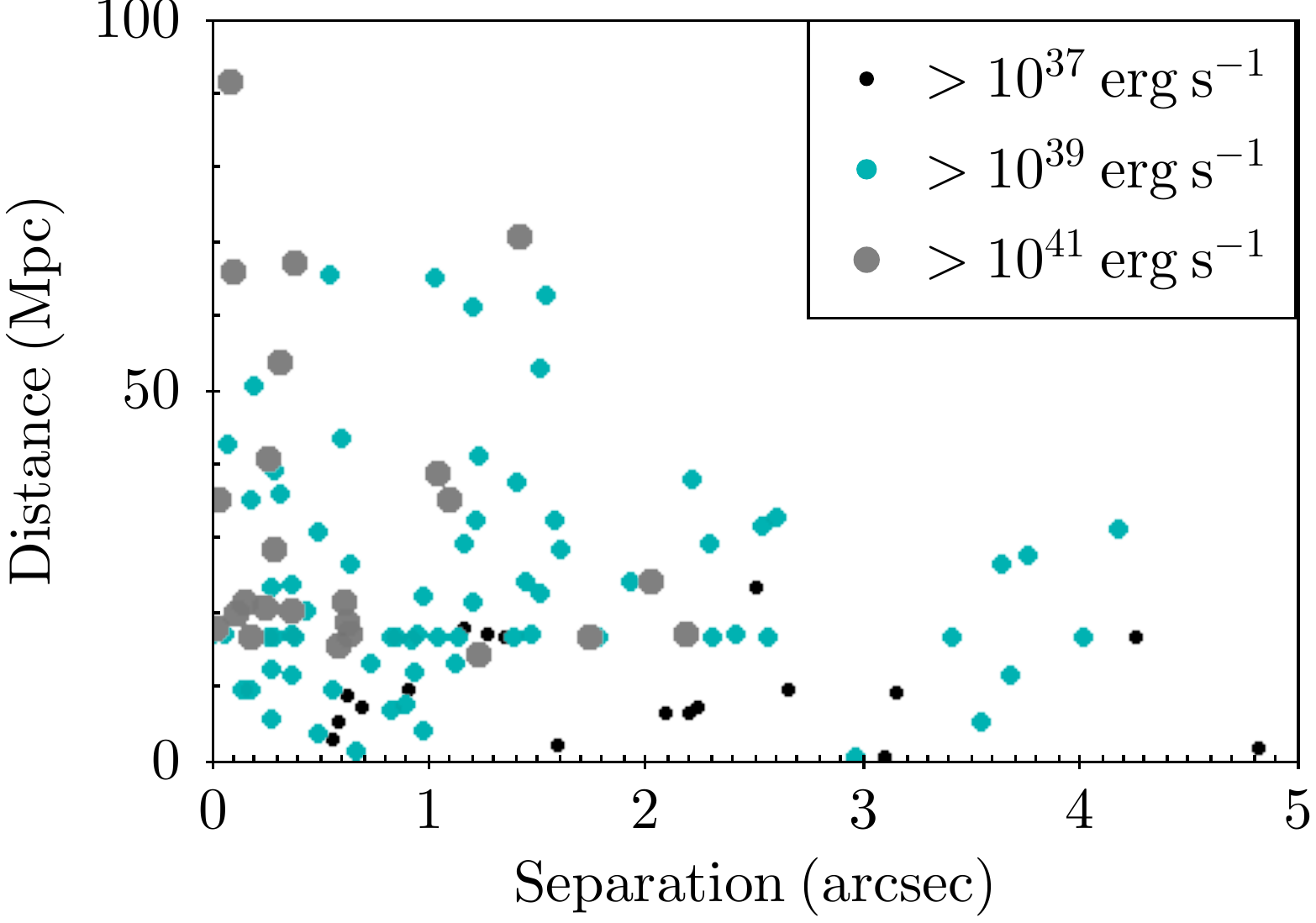}
\caption{Average X-ray luminosity of the nuclei in the matched sample at different distances and separation (in arcseconds and in parsecs) from the optical centre. NGC 2342 was excluded from the left plot, as the equivalent physical separation from the optical nucleus was $>$ 1000 pc.}
\label{fig:distsep}
\end{figure*}

\begin{figure*}[!ht]
    \centering
  
\includegraphics[width=0.4\textwidth]{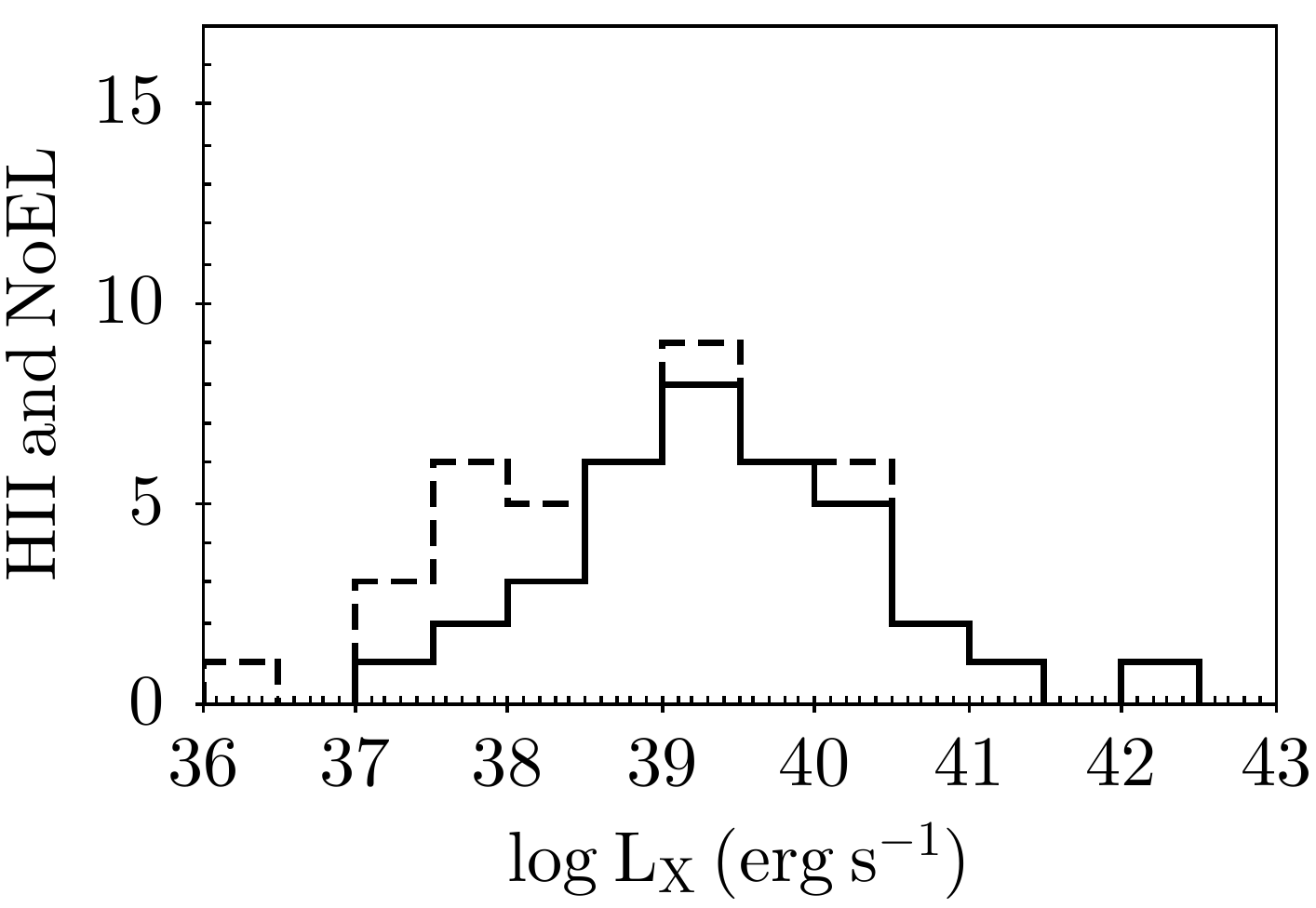}\label{fig:hii}
\includegraphics[width=0.4\textwidth]{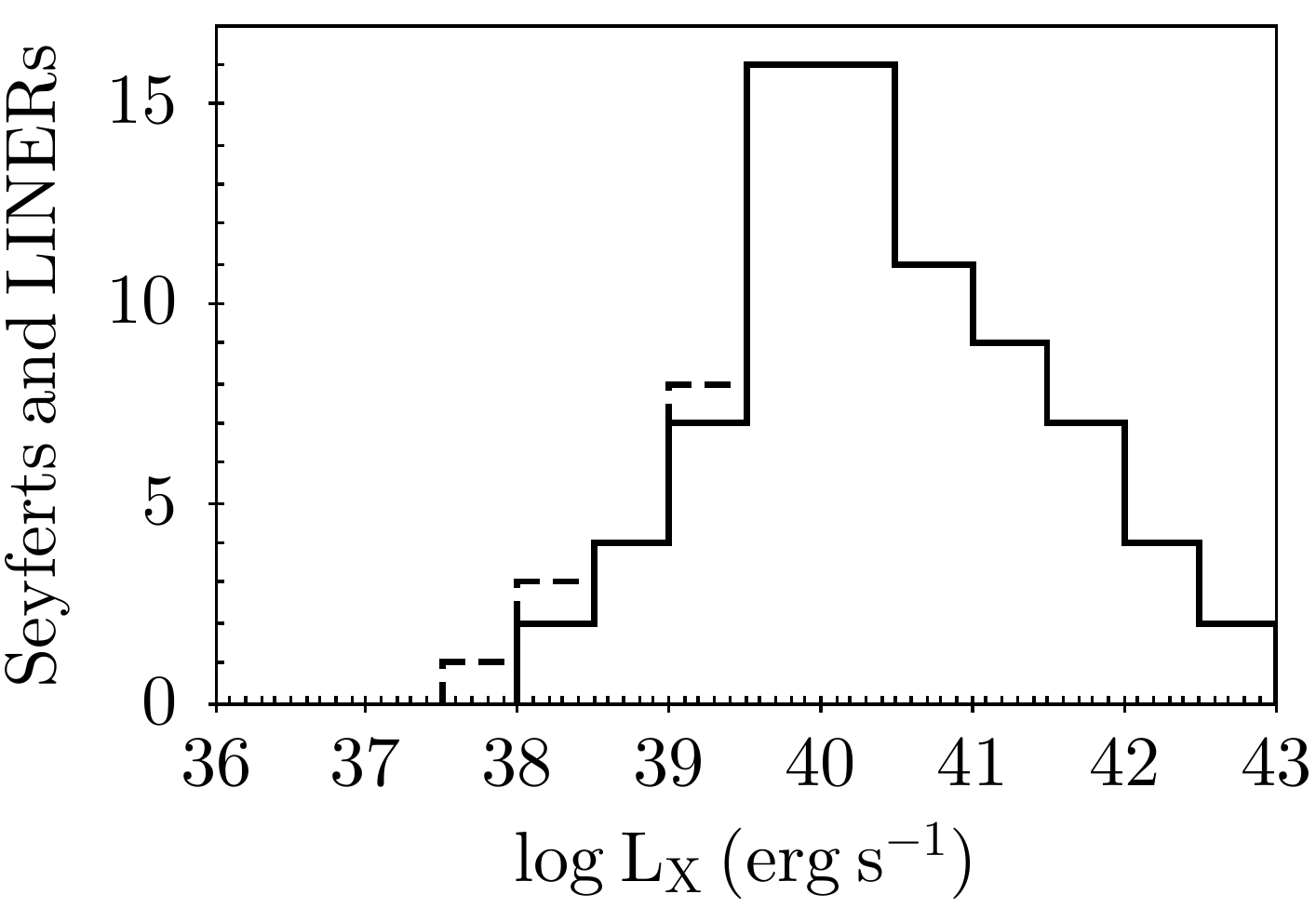}\label{fig:sl}

\caption{Histograms of the average nuclear X-ray luminosities of the galaxies within the selected sample (solid black line), for \hii and NoEL nuclei and for Seyfert and LINER nuclei. Flux upper limits  are included for non-detections - dashed lines.} 
\label{fig:lum}
\end{figure*}

None of the point sources in the selected nuclear sample have X-ray luminosities below $10^{37}$ \ergsec (Fig. \ref{fig:lum}). In Figure \ref{fig:distsep}, the  X-ray luminosity  is plotted at different host distances and separations from the optical centres.
In the right panel, the separation from the optical nucleus is given in arcseconds. This is converted to physical separation (in parsecs) in the left panel\footnote{Using the relation ${R}_{pc}={(R_{arcsec}} {D_{pc}} / 206265)$ and scaling for the nuclear region. Data for the semi-major axes was obtained from NED.}. The plots show that the nuclear sources with the lowest luminosity in our sample (between $\sim 10^{37}-10^{39}$ \ergsec) are found in the nearest host galaxies. As well as this, the majority of the most luminous sources 
 are found at small separations from the optical nucleus, as would be expected for LLAGN present in the sample. About half of all nuclei (58 out of 114) are found between 0 and 1 arcsec, and the average X-ray luminosities of 36 
of these exceed  10$^{39}$ \ergsec. In our sample, 66 nuclei are found within 100 pc from the optical position, and 
38 of them have  $L_{X} > 10^{39}$ \ergsec.

In the histograms presented in Fig. \ref{fig:lum}, the optically-defined \hii and NoEL nuclei exhibit lower X-ray luminosities than the Seyfert and LINER nuclei, as would be expected for generally agreeing optical and X-ray data, since the former should be less likely to have an active nucleus to contribute to their X-ray luminosity output.  Flux upper limits\footnote{http://www.ledas.ac.uk/flix/flix.html} are included for the non-detections. The average X-ray luminosity  for each optical nuclear subset is included in Table \ref{table:lum}. Confusion with X-ray binaries is expected for the sources with $L_{X} \lesssim 10^{39}$ \ergsec  \cite{zhang:2009}, since this is the range in which the luminosities of these sources and \llagns overlap.

\begin{figure*}[!t]
    \centering
 \includegraphics[width=0.35\textwidth]{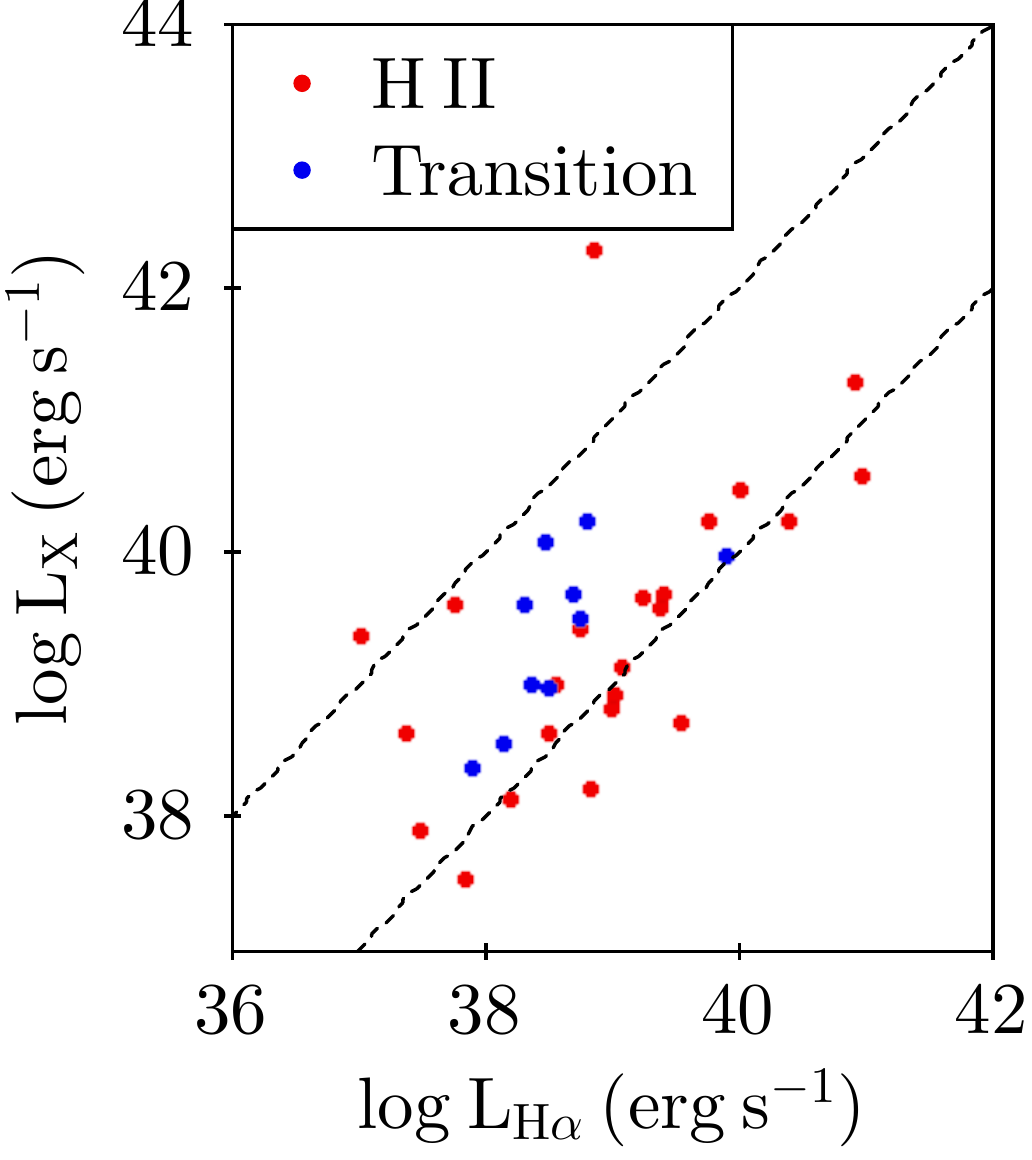}\label{fig:ha}
\includegraphics[width=0.35\textwidth]{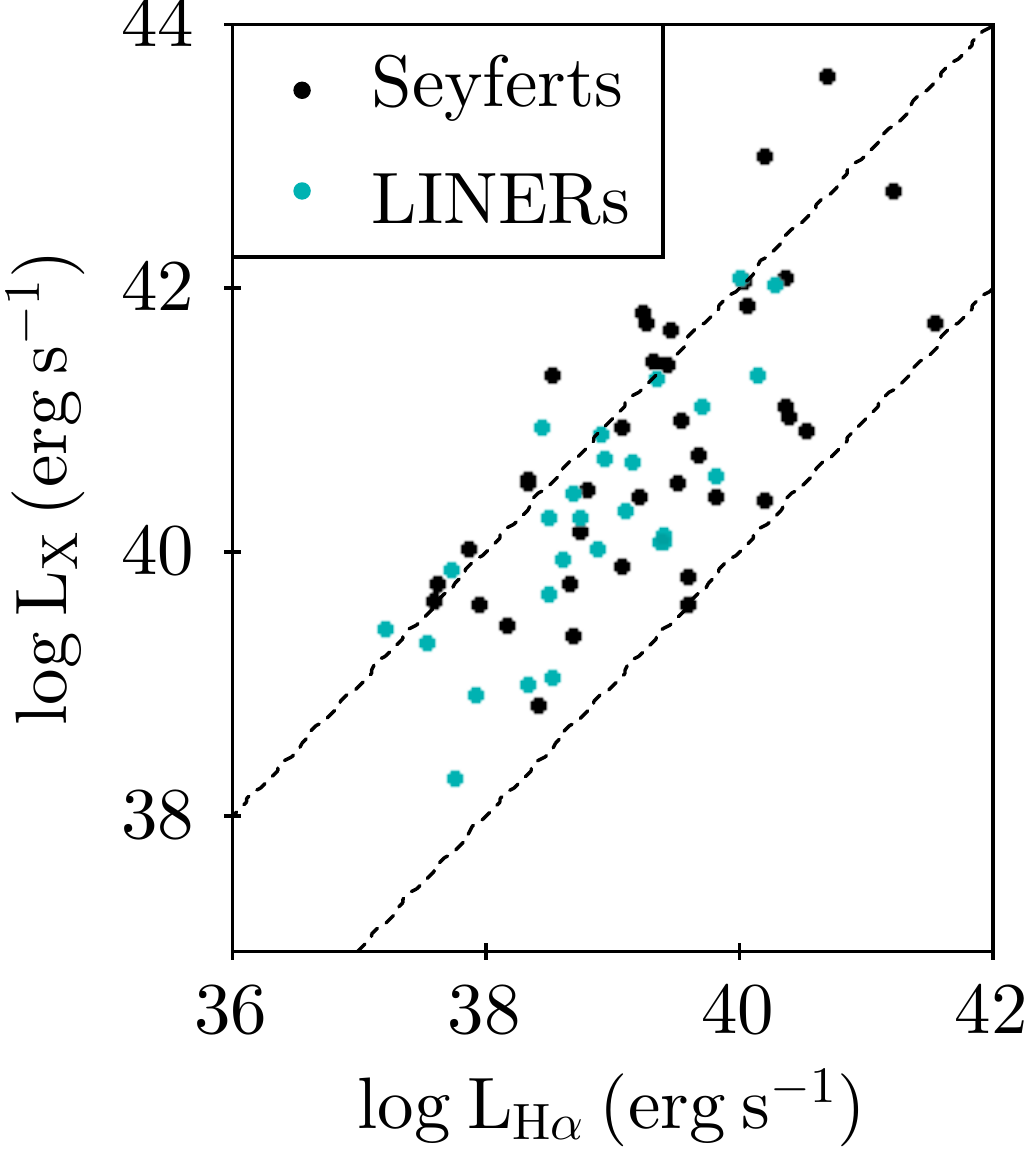}\label{fig:hal}
\caption{Correlation between $L_{H\alpha}$  and $L_{X}$ (0.2-10 keV) of the nuclei with point X-ray sources in our matched sample. The  range 100 $> L_{X}/L_{H\alpha} >$ 1  is plotted for optical Seyferts and LINERs (right panel), as well as for optical \hii and transition-LINERs (left panel).}
\label{fig:lum1}
\end{figure*}

The correlation between $L_{H\alpha}$ \cite{ho:1997a} and $L_{X}$ (0.2-10 keV) of the nuclei with point-like X-ray sources in our sample is plotted in Fig. \ref{fig:lum1}.  The typical for an (unobscured) accreting nuclear black hole range 100 $> L_{X}/L_{H\alpha} >$ 1 \cite{zhang:2009,roberts:2000,ho:2008} is plotted for optical Seyferts and LINERs, and compared to optical \hii and transition-LINERs. This is a good test for our sample, as for a star-forming nuclear cluster, $L_{X}/L_{H\alpha}$ is expected to be $ \lesssim$ 0.1 \cite{zhang:2009}.

\subsubsection{Black hole mass estimates and Eddington ratios}\label{bhm}

\begin{figure*}[!t]
    \centering
\includegraphics[width=0.5\textwidth]{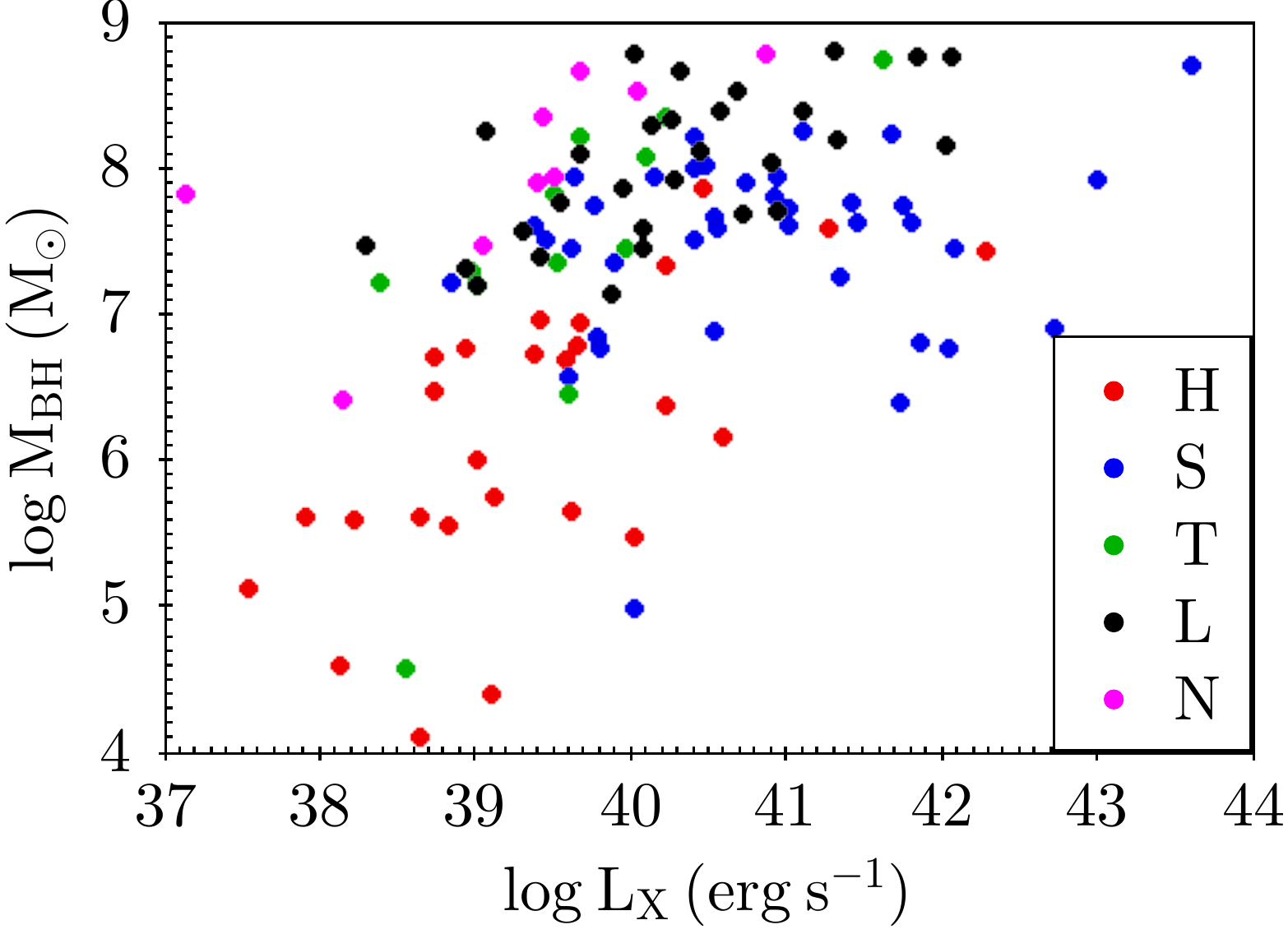}\label{fig:mbh}
\caption{Black hole mass estimates versus X-ray luminosity for the nuclear point sources. H stands for  \hii nuclei, S - Seyferts, L - pure LINERs, T - transition LINERs, N - NoELs.  }
\label{fig:bhmass}
\end{figure*}

The calculated from the black hole masses accretion rates serve as an approximation to the expected actual accretion rates, as a RIAF should be present in the LLAGN at small radii from the nucleus instead of an optically-thick, physically-thin accretion disc.  The plot of black hole mass, $M_{BH}$, against X-ray luminosity, $L_{X}$, for the different optical nuclear subsets, follows the expected trend for Seyfert and LINER galaxies to host SMBHs of higher mass (associated with higher X-ray luminosity) compared to, e.g., \hii nuclei (Fig. \ref{fig:bhmass}). However, optical transition nuclei cannot clearly be distinguished  from Seyferts and pure LINERs on this plot. The estimated average $M_{BH}$ for the Seyferts  (Table \ref{table:eddratio}) is, surprisingly, lower than the other (potentially) LLAGN-hosting subsets and the NoEL nuclei.

The ratios $\lambda_{Edd}$ for each subsample are less than $10^{-2}$ (sub-Eddington)\cite{exploring:2010, ho:2009a}, consistent with a radiatively inefficient accretion state (Table \ref{table:eddratio}). Interestingly, the value of $\lambda_{Edd}$ for the \hii subset is very close to the result for Seyfert nuclei, indicating that it is worth to look for LLAGN hidden in some of the the optically-classified \hii nuclei. The same applies to Transition LINERs, as the $\lambda_{Edd}$ result is very close to the result for pure LINERs. This suggests 
either very strong contamination or a possible disagreement between the optical and X-ray classification of these nuclei (see Section  \ref{sn:fitting}). 

\begin{table*}[!h]
\centering
\begin{tabular}{|lcc|}
\hline
\textbf{Nucleus type}& log($L_{Bol}/L_{Edd}$)& ${M}_{BH}(\Msun)$\\ 
\hline
\hline
Seyferts	& -3.04& (6.26$\pm 1.37)\times 10^{7}$\\
pure LINERs	& -4.46&(1.86$\pm 0.36)\times 10^{8}$\\
transition LINERs	& -4.40	&(1.02$\pm 0.45)\times 10^{8}$	\\
\hii	& -3.34&(8.55$\pm 3.12)\times 10^{6}$	\\
NoEL	& -6.28	 &(2.07$\pm 0.70)\times 10^{8}$\\
\hline
\hline
\textbf{All}& -3.81	&(1.13$\pm 0.14)\times 10^{8}$		\\
\hline
\end{tabular}
\caption{Average Eddington ratios and black hole mass estimates (from velocity dispersions) for each optically-defined nuclear subset in our sample.}
\label{table:eddratio}
\end{table*}

\subsection{Results of the spectral fitting for the X-ray nuclei of our \hii and transition-LINER optical subsets}\label{sn:fitting}

\subsubsection{Statistics of the spectral sample}
Excluding the nuclear sources flagged as extended in \3xmm  left us with a total of 12 transition-LINERs and 26 \hii nuclei (table \ref{table:lum}). Unfortunately, no spectral products were available for the nuclei of two of the transition galaxies (NGC 3953 and NGC 5879), as well as for one of the \hii galaxies (NGC 3877). Further four \hii galaxies had far too few counts for the central source to allow a meaningful fit (NGC 3077, NGC 3184, NGC 3319, NGC 5775).
Galaxies with nuclear X-ray sources also included in other studies (e.g. NGC 4321, NGC 4449 \cite{roberts:2000}) had only extended detections in the \3xmm catalogue, and were not considered as LLAGN candidates in this project. 

\subsubsection{The best-fit model} \label{best}
In Table 7, the results of the spectral fits are presented. The upper part of the table shows the initial parameters (the redshift $z$, the galactic column density $N_{H,Gal}$) and the fitted model parameters (the host column density $N_{H,int}$, the temperature $k_{B}T$, the photon index $\Gamma$, the model luminosities) for the transition galaxies with detected X-ray point nuclear sources. The lower table includes the initial parameters  and results for the \hii galaxies with detected X-ray point nuclear source. Errors  within a 90\% Gaussian uncertainty range for the physical parameters are included in the table. In the case of multiple observations of the same nuclei, the spectral fits are shown in chronological order. 

The value of $\chisq$ and the number of degrees of freedom ($\nu$) are shown in each row for the absorbed power-law  with thermal plasma model (\texttt{const$\ast$tbabs$\ast$ztbabs$\ast$(apec+po)}), which successfully fitted the majority of the spectral datasets. The improvement over the simple  absorbed power-law (\texttt{const$\ast$tbabs$\ast$ztbabs$\ast$po}) is shown as $\Delta \chisq$/$\Delta \nu$.

\begin{figure*}[!t]
 \centering
\includegraphics[width=0.43\textwidth, height=5.1cm]{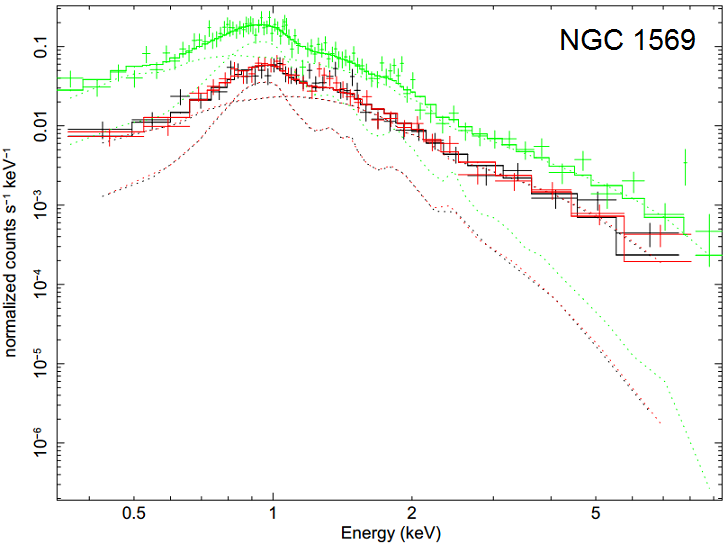}
\includegraphics[width=0.45\textwidth]{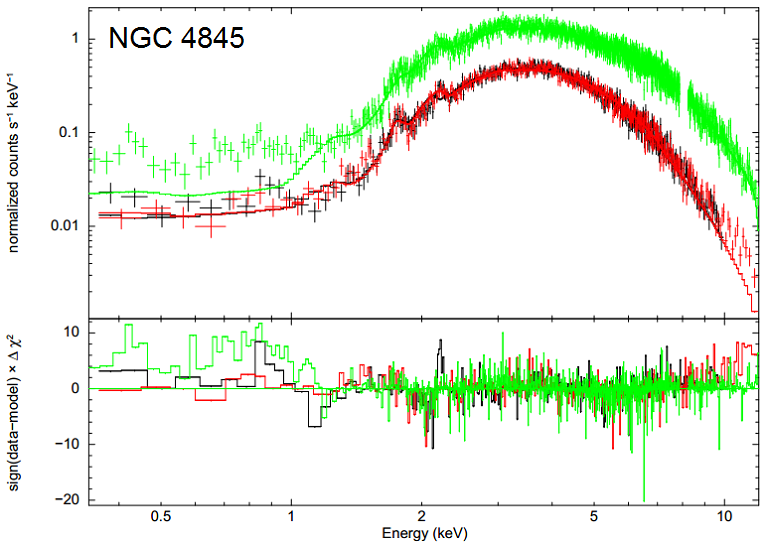}
\includegraphics[width=0.452\textwidth, height=5.3cm]{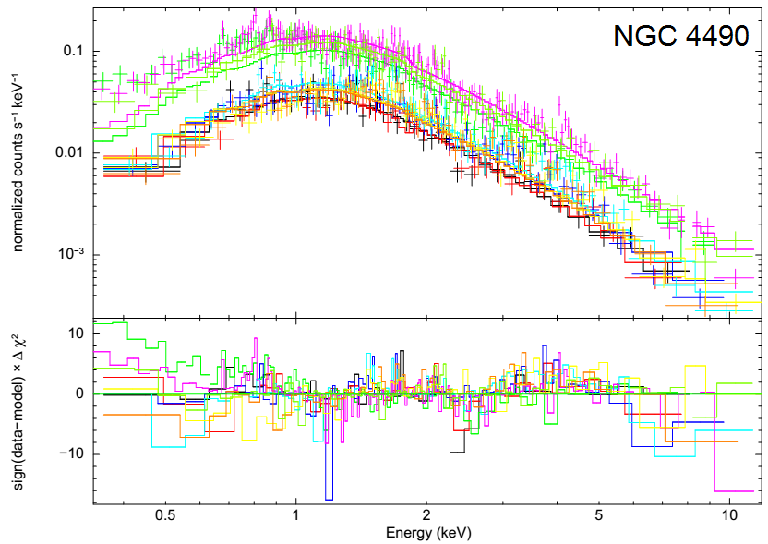}
\includegraphics[width=0.45\textwidth, height=5.25cm]{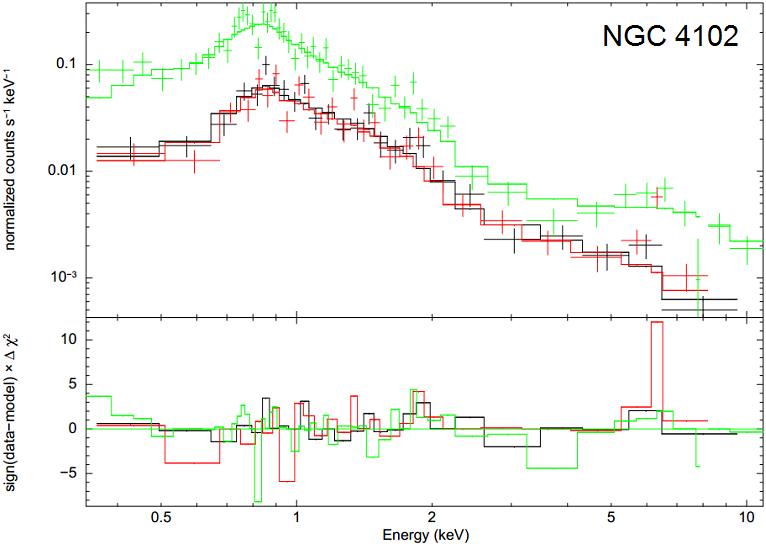}
\caption{Example fitted spectra in our sample (pn, MOS1, MOS2):  NGC 4102  has a transition-LINER spectrum; NGC 4845 - a Seyfert 2 spectrum; NGC 1569 - a starburst spectrum (with shown model components). The offset between the pn and MOS spectra is due to differences in flux calibration. The spectrum of NGC 4102 could contain a Fe K$\alpha$ line at 6.4 keV, but too few datapoints are found at this energy. The plot for NGC 4490 shows a simultaneous fit  from 3 different observations and the resulting strong residuals, as all sets could not be fitted with the same flux.}
\label{fig:spectra}
\end{figure*}

 In almost all cases, $\rchi$ $\approx$ 1 was obtained for the \texttt{(apec+po)}-based model, and thus it was possible to obtain model parameters with errors and flux estimates for the different spectral components. The only exception to this was NGC 4102, where two power-law components were introduced to obtain an acceptable fit (Fig. \ref{fig:spectra}, compare to Fig. \ref{fig:vs}). None of the fitted datasets showed a preference towards the simple absorbed \texttt{po} model. Overall, there was no statistically significant improvement in $\chisq$ of the \texttt{(apec+po)}-based model over the \texttt{po}-based model in $\sim$ 23\% of the Transition nuclei fits and $\sim$ 24\% of the \hii nuclei fits. 
Therefore, in those cases adding the thermal component was not essential for the fit.  

Overall, the absorbed \texttt{(apec+po)} model lead to good statistics and a reasonable spread of physical parameters, and was therefore a good representation of our data. The \texttt{apec} model temperatures agree with the expected values of $\sim$ $ 10^{6}-10^{7}$ K, with lower values for NGC 5746, NGC 598, NGC 4845 ($\sim 5-6 \times 10^{5} $ K). However, the null hypothesis of the \texttt{(apec+po)}-based model fit was questioned or rejected in several cases. For NGC 520 and NGC 4298, this was due to noisy data and a total number of counts very close to the minimum for a meaningful fit. In some cases (NGC 598, NGC 2903, NGC 4517), despite obtaining $\rchi$ $\sim$ 1,  the data are complex, and require more model components  to improve the statistics. 

Overall, the typical shape of our sample spectra was similar to either NGC 1569 (consistent with thermal plasma with unresolved point-like sources, e.g. X-ray binaries) or NGC 4490 (a LLAGN candidate)  (Fig. \ref{fig:spectra}, also compare to Fig. \ref{fig:avg} and \ref{fig:vs}), as expected.  
The spectrum of NGC 4845 (a known Seyfert 2 \cite{veron:2006}) stood out in our sample as the only strongly absorbed at lower energies. The residual in the low energy region ($\lesssim$ 1.5 keV) caused by relatively poor quality data, lead to the rejection of the model fit, despite a very good agreement with the model at higher energies (Fig. \ref{fig:spectra}).  

The contributions to the obtained unabsorbed source luminosity, $L_{intrinsic}$, are also included in the table. The power-law contribution to $L_{intrinsic}$ and its error are presented as percentage of the obtained intrinsic luminosity. 

\subsubsection{Variability check and Fe K$\alpha$ line search?}
After simultaneously fitting all available datasets for the nuclei with multiple observations, the  variations in the physical parameters  appear to be due to intrinsic variability from the flux in the case of NGC 4490 and NGC 6217, rather than from the model parameters. This is consistent with flux variability on a timescale of days between the observations. 
In the case of NGC 6217, the  result of  $\Gamma > 3$ (outside the accepted range, Section \ref{fitting}) and large errors for the two fits are likely to be due to noisy data and a relatively low number of counts available in the two datasets. 

No Fe K$\alpha$ line was confirmed in any of the fits. A feature resembling a Fe K$\alpha$ line (at $\sim$ 6.4 keV) is noticeable only in the MOS1 fit residual of NGC 4102 (Fig. \ref{fig:spectra}), but  adding a \texttt{gauss} component to our model still lead to an equivalent width limit of $<$ 1 keV. There are too few data points at this energy for this to be a conclusive test for NGC 4102. 

\subsubsection{The LLAGN candidates}\label{cand}
 
Some of the fitted nuclei have an intrinsic luminosity strongly dominated  by the thermal contribution (NGC 410, NGC 5846, NGC 4517), but the majority of the nuclei have  $> 50\%$ of their estimated luminosity output dominated by the power-law  component.  Clearly,  this does not automatically enlist all those nuclei as potential LLAGN. Using the method described in Section \ref{checks} lead to the final determination of our LLAGN candidates. Our simple image analysis test with the available \textit{Chandra} data indicates that the nuclei in the LLAGN-candidate sample are either consistent with a single point source, or with a single point source embedded in extended emission, with an average FWHM(Nucleus)-FWHM(PSF) of $\sim$ 0.2-0.3". There is disagreement between our simple check and literature regarding the extent of the nucleus of NGC 660, which was found to be more consistent with an extended source in two \textit{Chandra} studies \cite{zhang:2009, liu:2011}.  As well as this, the classification was inconsistent between the two studies for some of the nuclei, as can be seen in Table \ref{table:candidates}.  Based on our analysis, 40\% the transition galaxies and $\approx$ 43\% of the \hii galaxies in our sample might host a LLAGN. 

Other details for our LLAGN candidates  presented in Table \ref{table:candidates} are the host galaxy Hubble type (T) \cite{ho:1997a},  the offset of the source from which our processed \xmm spectrum was extracted ($\delta$), the black hole mass estimate from galaxy velocity dispersions (log$M_{BH}$, see section \ref{mass}), and the Eddington ratios (log$\lambda_{Edd}$), estimated from the obtained intrinsic luminosity via Eqs. \ref{eq:Lbol} and \ref{eq:Eddratio}. 
All of the LLAGN candidates are spiral galaxies, with 8 early-types (T from 0 to 4) and 5 late-types  (T from 5 to 10). 
 Out of the 13 candidates (comprising $\approx$ 42\% of our fitted sample),  a literature search showed that at least 5 have been classified as possible accretion-powered nuclei based on X-ray image and spectral analysis of \textit{Chandra} and/or \xmm data \cite{zhang:2009, liu:2011, martin:2011}.  Further 3 of the nuclei are marked in NED as Seyfert-2 types \cite{veron:2006}.  However, as noted in  section \ref{best}, only the EPIC spectrum of NGC 4845 was distinctive and in obvious agreement with the Seyfert-2 classification. 
Our  $\lambda_{Edd}$ estimates are sub-Eddington; for two of the galaxies (NGC 4845 and NGC 4490) log$\lambda_{Edd}$ is $>$ -2.0. As the Eddington ratio is a way to quantify accretion, this is again in agreement with the classification of NGC 4845 and suggestive of nuclear activity in NGC 4490. 
All candidates lie within $\sim$ 200 pc from the optical nuclear position, except for NGC 2342, where the source is found at more than 1000 pc from the optical nucleus.  The NGC 2342 fitted model, obtained spectral slope, temperature and luminosity are consistent with an \xmm study of the galaxy \cite{roberts:2005}. As investigated in detail in the article,  this galaxy (in the galaxy pair NGC 2342/2341) is more likely to have an integrated spectrum from X-ray binaries,  typical for merger systems \cite{roberts:2005}, rather than a LLAGN. These results are discussed in Section \ref{discussion}.

\begin{table*}[!hb]
\centering
\begin{tabular}{|lcccccccc|}
\hline
Name& T & $\delta$ (") & $\delta$ (pc) & Core &$N_{r}$& log$M_{BH}$  & log$\lambda_{Edd}$ & Notes\\ 
\hline
\hline
NGC 660 &1 & 3.68 & 210.4 & III$^{a,b}$ & 1 & 7.4 & -4.4 & \hii/L$^{b}$\\
NGC 3627 & 3& 2.09 & 66.8 & I$^{a}$/III$^{b}$ & 1 & 7.3 & -4.8 & L/Sy2$^{b}$\\
NGC 4713 &7 & 1.17 & 101.3 & I$^{a}$ & 1 & 4.6 & -2.2 &  \hii/L$^{b, c}$\\
NGC 5055 &4 &0.69 & 24.1 & II$^{a}$/I$^{b}$ & 1 & 7.2 & -4.7 & \hii/L$^{b}$\\
\hline
\hline
IC 342 &6 & 0.56 & 8.1 & I$^{a}$ &  & 6.5 & -4.3 &  \hii$^{b,c}$\\ 
NGC 2342 &0 & 3.49 & 1174.6 &  &  & 7.6 & -3.1 & \hii$^{b}$ \\ 
NGC 4102 &3 &0.37 & 30.8 & I$^{b}$ & 1 & 7.9 & -4.1 & \hii /L$^{b,c}$ \\
NGC 4298 &5 & 1.35 & 110.2 &  &  & 5.5 & -3.3 &  \hii$^{b}$\\ 
 NGC 4490 &7& 0.89 & 33.8 & IV$^{a}$/I & 1 & 5.7 & -1.2 & \hii$^{c}$\\ 
NGC 4654 &6 & 0.82 & 66.8 & II$^{b}$  &  1 & 5.7 & -3.3 & \hii$^{c}$\\ 
NGC 4845 &2 &0.59 & 44.7 &  &  & 7.4 & -1.5 & Sy2$^{c}$\\
NGC 5248 &4 &1.52 & 167.1 &  &  & 6.9 & -3.9 & \hii/Sy2$^{c}$ \\
NGC 6217 &4& 0.38 & 43.5 &  & & 6.4 & -2.8 &\hii/Sy2$^{c}$ \\ 
\hline
\end{tabular}
\caption{Our LLAGN candidates. 
The \3xmm source offset from the optical nucleus is given in arcsec and pc. The Eddington ratios were calculated using the fitted intrinsic luminosities.  $M_{BH}$ is in solar masses. L stands for LINER nucleus, Sy 2 for Seyfert 2. The number of resolved sources ($N_{r}$) by \textit{Chandra} within 6" is given. I -  dominant point source at the nuclear position; II - point source and extended emission; III - extended source only; IV - no source at nuclear position;  $^{a}$ \textit{Chandra}  X-ray core was analysed in \cite{zhang:2009};   
$^{b}$ \textit{Chandra}  X-ray core analysed in \cite{liu:2011}; $^{c}$ NED classification.}
\label{table:candidates}
\end{table*}

\begin{landscape}
Table 7: Results of the spectral fitting of the \xmm observations. Upper table - transition galaxies with detected X-ray point nuclear source. Lower table - \hii galaxies with detected X-ray point nuclear source. Power-law dominated spectra (contribution to $L_{intrinsic} >$ 82\%) could potentially indicate the presence of an active nucleus.

\begin{longtable}{|ccccccccccc|}

\hline
\textbf{Transition}&$z$& $N_{H,Gal}^{a}$	& $N_{H,int}^{b}$	& $k_{B}T$ (keV)	& $\Gamma$	&
$\chisq$/$\nu$	& $\Delta \chisq$/$\Delta \nu^{c}$      & $L_{absorbed}^{d}$& $L_{intrinsic}^{d}$& $L_{p. law}(\%L_{intr})$\\
\hline
\hline
NGC 410	&0.0177& 5.4 & 0.05$^{+0.02}_{-0.02}$ & 0.86$^{+0.02}_{-0.01}$ &1.70$^{+0.42}_{-0.45}$ & 	 287.8/277&1057.4/2  &  3.91$^{+0.11}_{-0.12}$$\times 10^{41}$ &5.07$^{+0.24}_{-0.26}$$\times 10^{41}$ &12.10$^{+2.68}_{-2.90}$ $\%$\\
 
...&...&...&$<$0.03&0.85$^{+0.02}_{-0.02}$ &1.22$^{+1.04}_{-1.52}$ &171.6/181 &569.1/2 &  3.98$^{+0.27}_{-0.22}$$\times 10^{41}$ &5.36$^{+0.45}_{-0.32}$$\times 10^{41}$ &9.93$^{+3.56}_{-4.55}$ $\%$\\
	
NGC 660	& 0.0028 & 4.9 &0.25$^{+0.36}_{-0.07}$ &0.72$^{+0.19}_{-0.56}$&1.89$^{+0.29}_{-0.20}$&62.12/44&14.3/2  
 & 4.38$^{+0.41}_{-0.43}$$\times 10^{39}$ &7.52$^{+9.20}_{-1.36}$$\times 10^{39}$ &82.11$^{+16.57}_{-8.59}$ $\%$\\ 

NGC 3627&0.0024& 2.4 & $<$0.82 & 0.49$^{+0.13}_{-0.12}$ &1.90$^{+0.11}_{-0.12}$&81.3/74&59.0/2  
&2.21$^{+0.15}_{-0.13}$$\times 10^{39}$ & 2.55$^{+0.15}_{-0.15}$$\times 10^{39}$ &85.70$^{+5.84}_{-6.23}$ $\%$\\  

NGC 4459 &0.0040& 2.7&$<$0.11&0.41$^{+0.18}_{-0.09}$&1.90$^{+0.21}_{-0.20}$&55.15/58& 47.5/2& 4.39$^{+0.40}_{-0.42}$$\times 10^{39}$ &5.72$^{+0.73}_{-0.64}$$\times 10^{39}$ &81.30$^{+8.38}_{-9.17}$ $\%$\\ 

NGC 4569	&-0.0008& 2.5 & $<$0.01  &0.76$^{+0.02}_{-0.02}$&1.93$^{+0.06}_{-0.05}$&335.2/320&526.2/2 
 &8.61$^{+0.24}_{-0.21}$$\times 10^{39}$ &9.75$^{+0.26}_{-0.26}$$\times 10^{40}$ &70.88$^{+3.20}_{-3.18}$ $\%$\\

NGC 4713&0.0022& 2.0&$<$0.07&0.25$^{+0.04}_{-0.03}$&1.99$^{+0.23}_{-0.21}$&80.2/81&42.84/2 
&1.69$^{+0.16}_{-0.17}$$\times 10^{39}$ &1.94$^{+2.17}_{-2.02}$$\times 10^{39}$  &76.33$^{+6.92}_{-7.19}$ $\%$\\

NGC 5055&0.0016& 1.3&$<$0.82&0.35$^{+0.05}_{-0.03}$&1.42$^{+0.25}_{-0.26}$&46.6/42&80.3/2  
&2.55$^{+0.33}_{-0.33}$$\times 10^{39}$ &2.37$^{+0.30}_{-0.29}$$\times 10^{39}$ &74.97$^{+10.58}_{-10.61}$ $\%$\\

NGC 5354$^{e}$ &0.0086& 1.0 &$<$0.12&0.89$^{+0.15}_{-0.20}$ &1.67$^{+0.46}_{-0.21}$&13.4/14&14.3/2  
&8.39$^{+1.05}_{-1.61}$$\times 10^{39}$ &8.62$^{+1.31}_{-1.16}$$\times 10^{39}$ &80.72$^{+16.07}_{-17.34}$ $\%$\\

NGC 5746	&0.0057& 3.3 &0.13$^{+0.08}_{-0.08}$&0.05$^{+0.03}_{-0.01}$ &1.33$^{+0.10}_{-0.10}$& 158.7/135&7.1/2
&1.82$^{+0.10}_{-0.10}$$\times 10^{40}$ &3.05$^{+2.20}_{-1.09}$$\times 10^{40}$ &67.20$^{+3.27}_{-3.40}$ $\%$\\

...&...&...&0.25$^{+0.06}_{-0.06}$&0.03$^{+0.01}_{-0.01}$ &1.43$^{+0.07}_{-0.07}$ &249.4/234&23.9/2 
&2.15$^{+0.09}_{-0.09}$$\times 10^{40}$ &6.54$^{+5.31}_{-3.98}$$\times 10^{40}$ &37.40$^{+1.34}_{-1.37}$ $\%$\\

...&...&... &0.29$^{+0.07}_{-0.07}$ &0.04$^{+0.01}_{-0.01}$&1.58$^{+0.10}_{-0.09}$&197.4/164&26.7/2
&1.61$^{+0.08}_{-0.08}$$\times 10^{40}$ &9.47$^{+10.34}_{-7.88}$$\times 10^{40}$ &22.71$^{+1.05}_{-1.06}$ $\%$\\

NGC 5846	&0.0057& 4.3 &$<$0.01&0.74$^{+0.01}_{-0.01}$&2.19$^{+0.21}_{-0.25}$
&416.5/355& 1471.2/2
&1.32$^{+0.02}_{-0.02}$$\times 10^{41}$ &1.60$^{+0.02}_{-0.03}$$\times 10^{41}$ & 11.89$^{+1.91}_{-2.28}$ $\%$\\
\hline
\hline

\textbf{\hii}&$z$& $N_{H,Gal}^{a}$	& $N_{H,int}^{b}$	& $k_{B}T$ (keV)	& $\Gamma$	&
$\chisq$/d.o.f.	& $\Delta \chisq$/$\Delta \nu^{c}$      & $L_{absorbed}^{d}$& $L_{intrinsic}^{d}$& $L_{p. law}(\%L_{intr})$\\
\hline
\hline
IC 342 &0.0001&30.2&0.17$^{+0.06}_{-0.05}$&0.78$^{+0.13}_{-0.08}$&2.12$^{+0.13}_{-0.13}$
&110.5/119& 57.3/2
&8.32$^{+0.47}_{-0.49}$\a &1.66$^{+0.14}_{-0.13}$$\times 10^{39}$& 82.45$^{+7.55}_{-6.90}$ $\%$\\

...&...&...&0.15$^{+0.14}_{-0.05}$&0.73$^{+0.04}_{-0.04}$
&2.23$^{+0.15}_{-0.15}$&187.8/154& 243.6/2
&4.61$^{+0.24}_{-0.24}$\a &1.13$^{+0.26}_{-0.09}$$\times 10^{39}$ &60.77$^{+6.98}_{-6.45}$ $\%$\\

NGC 520$^{e}$
&0.0076&3.3&$<$0.13&0.97$^{+0.30}_{-0.30}$&1.30$^{+0.31}_{-0.28}$&10.5/20$^{\dagger}$& 12.3/2 &1.31$^{+0.20}_{-0.22}$$\times 10^{40}$ &1.53$^{+0.19}_{-0.19}$$\times 10^{40}$ & 89.60$^{+10.40}_{-13.29}$ $\%$\\ 

NGC 598&-0.0006&5.6&0.38$^{+0.01}_{-0.01}$&0.04$^{+0.01}_{-0.01}$&2.40$^{+0.02}_{-0.02}$& 1601.3/1109$^{\star}$&403.9/2
&1.02$^{+0.01}_{-0.01}$$\times 10^{39}$ &3.43$^{+0.34}_{-0.32}$$\times 10^{39}$ & 63.04$^{+1.35}_{-1.34}$ $\%$\\

NGC 1569 &-0.0003 & 22.4 &0.10$^{+0.04}_{-0.04}$&0.95$^{+0.05}_{-0.06}$&2.41$^{+0.18}_{-0.17}$& 159.1/158&111/2
&1.26$^{+0.06}_{-0.06}$\a &2.47$^{+0.28}_{-0.24}$\a & 75.79$^{+11.97}_{-10.53}$ $\%$\\

...&...&...&0.11$^{+0.04}_{-0.04}$&0.96$^{+0.05}_{-0.06}$
&2.43$^{+0.18}_{-0.17}$&153.8/157& 108.3/2
&1.25$^{+0.06}_{-0.06}$\a &2.48$^{+0.32}_{-0.28}$\a & 76.03$^{+12.18}_{-10.71}$ $\%$\\

NGC 2146& 0.0030& 7.3&0.15$^{+0.03}_{-0.03}$&0.75$^{+0.04}_{-0.04}$&1.60$^{+0.07}_{-0.07}$&306.8/241& 332/2
&5.47$^{+0.20}_{-0.21}$$\times 10^{40}$ &6.90$^{+0.25}_{-0.24}$$\times 10^{40}$ & 80.32$^{+3.04}_{-3.14}$ $\%$\\

NGC 2342&0.0176 &8.9 &0.08$^{+0.03}_{-0.03}$&0.79$^{+0.09}_{-0.09}$&1.94$^{+0.11}_{-0.11}$&166/165&52.2/2 &1.67$^{+0.08}_{-0.09}$$\times 10^{41}$ &2.19$^{+0.11}_{-0.10}$$\times 10^{41}$ &89.80$^{+5.44}_{-5.31}$ $\%$  \\

NGC 2903& 0.0018& 3.1&$<$0.01&0.64$^{+0.02}_{-0.02}$&2.24$^{+0.07}_{-0.04}$& 757.5/494$^{\star}$&1184.1/2&1.93$^{+0.03}_{-0.03}$$\times 10^{39}$ &1.87$^{+0.06}_{-0.03}$$\times 10^{39}$ &68.36$^{+3.01}_{-2.25}$ $\%$\\

NGC 3367&0.0101&2.9&$<$0.02&0.72$^{+0.05}_{-0.10}$&1.97$^{+0.12}_{-0.12}$&103.4/87&113.7/2 &5.23$^{+0.30}_{-0.31}$$\times 10^{40}$ &5.72$^{+0.52}_{-0.55}$$\times 10^{40}$ &74.20$^{+8.91}_{-7.50}$ $\%$ \\

NGC 4102$^{f}$ &0.0028&1.8&0.26$^{+0.08}_{-0.07}$&0.69$^{+0.08}_{-0.06}$&2.61$^{+0.43}_{-0.37}$& 142.3/96&162.6/4
&3.00$^{+0.21}_{-0.34}$$\times 10^{40}$ &4.74$^{+0.84}_{-0.80}$$\times 10^{40}$ &86.44$^{+13.56}_{-16.17}$ $\%$\\

NGC 4298 &0.0038&2.7&$<$0.53&0.43$^{+0.93}_{-0.25}$&1.58$^{+0.31}_{-0.28}$&11.2/17$^{\dagger}$& 3.2/2&1.33$^{+0.21}_{-0.25}$$\times 10^{39}$ &1.53$^{+2.36}_{-0.21}$$\times 10^{39}$ &93.93$^{+6.07}_{-13.38}$ $\%$ \\

\hline

NGC 4470& 0.0078&1.6&$<$0.23&0.32$^{+0.32}_{-0.22}$&2.39$^{+0.66}_{-0.42}$&40.9/31&3.6/2 &3.99$^{+0.63}_{-0.67}$$\times 10^{39}$ &4.60$^{+2.12}_{-0.89}$$\times 10^{39}$ & 87.88$^{+12.12}_{-19.43}$ $\%$ \\

NGC 4490& 0.0019&1.8 &0.28$^{+0.06}_{-0.03}$&0.62$^{+0.18}_{-0.30}$&1.91$^{+0.08}_{-0.08}$& 163.4/189&16.6/2&4.90$^{+0.20}_{-0.21}$$\times 10^{39}$ &7.15$^{+0.61}_{-0.32}$$\times 10^{39}$ &94.89$^{+5.11}_{-4.42}$ $\%$ \\

...&... &... &0.73$^{+0.10}_{-0.11}$&0.19$^{+0.02}_{-0.01}$&1.88$^{+0.06}_{-0.07}$&385.7/ 343 &41.8/2 &6.58$^{+0.19}_{-0.19}$$\times 10^{39}$ &2.54$^{+0.91}_{-0.89}$$\times 10^{40}$ &46.83$^{+2.83}_{-2.97}$ $\%$  \\

...&... &... &1.08$^{+0.15}_{-0.14}$&0.11$^{+0.01}_{-0.02}$&2.56$^{+0.12}_{-0.09}$&257.4/206 &23/2 &5.74$^{+0.23}_{-0.24}$$\times 10^{39}$ &2.09$^{+2.67}_{-1.47}$$\times 10^{40}$ &8.86$^{+1.33}_{-0.94}$ $\%$  \\

all$^{g}$&... &... &0.82$^{+0.07}_{-0.08}$&0.14$^{+0.03}_{-0.01}$&2.66$^{+0.05}_{-0.05}$&1264.8/747$^{\star}$ &39.5/2&| &| &|  \\

NGC 4517& 0.0038&1.9 &1.51$^{+0.09}_{-0.09}$&0.07$^{+0.01}_{-0.01}$&2.12$^{+0.06}_{-0.05}$&738.4/603$^{\dagger}$& 144.2/2&7.33$^{+0.15}_{-0.16}$$\times 10^{39}$ &1.98$^{+1.06}_{-1.33}$$\times 10^{42}$ & 0.87$^{+0.05}_{-0.05}$ $\%$\\

NGC 4536& 0.0060&1.8 &0.13$^{+0.07}_{-0.04}$&0.71$^{+0.05}_{-0.12}$&2.26$^{+0.24}_{-0.22}$&183.1/117$^{\star}$ & 143.7/2&3.47$^{+0.24}_{-0.25}$$\times 10^{39}$ &5.55$^{+1.03}_{-0.56}$$\times 10^{39}$ & 66.37$^{+15.02}_{-9.52}$ $\%$ \\

NGC 4647& 0.0047&2.2 & $<$0.12 &0.29$^{+0.04}_{-0.04}$&2.80$^{+0.86}_{-0.49}$&42.3/33&21/2 &8.19$^{+1.04}_{-1.10}$\a &1.29$^{+0.70}_{-0.19}$$\times 10^{39}$ &62.49$^{+37.52}_{-20.20}$ $\%$  \\

NGC 4654& 0.0035& 2.3& $<$0.09 &0.54$^{+0.17}_{-0.25}$&2.09$^{+0.43}_{-0.25}$&19.9/22&10.8/2&1.93$^{+0.20}_{-0.33}$$\times 10^{39}$ &2.18$^{+0.47}_{-0.24}$$\times 10^{39}$ &86.28$^{+11.03}_{-20.19}$ $\%$  \\

NGC 4845&0.0041 &1.6 & 10.37$^{+0.13}_{-0.12}$ &0.05$^{+0.90}_{-0.01}$&2.10$^{+0.02}_{-0.02}$& 2987.3/2396$^{\star}$ & 104.9/2
&1.66$^{+0.01}_{-0.01}$\e &6.27$^{+0.20}_{-0.19}$\e &100.00$^{+0.00}_{-3.11}$ $\%$  \\

NGC 5248& 0.0038& 2.1&$<$0.10 &0.53$^{+0.24}_{-0.25}$&1.85$^{+0.36}_{-0.28}$&49.9/35& 41.3/2&8.68$^{+0.83}_{-}$$\times 10^{39}$ &9.64$^{+0.91}_{-0.89}$$\times 10^{39}$ &77.48$^{+10.60}_{-12.14}$ $\%$  \\

NGC 5457
&0.0008&1.2 &$<$0.09&0.24$^{+0.31}_{-0.13}$&1.88$^{+0.42}_{-0.36}$&2.7/10$^{\dagger}$&3.5/2
&2.36$^{+0.36}_{-0.44}$\a &2.49$^{+0.41}_{-0.40}$\a &86.38$^{+13.18}_{-15.14}$ $\%$  \\

NGC 6217& 0.0045& 4.0&0.29$^{+0.16}_{-0.10}$ &0.64$^{+0.11}_{-0.06}$&4.21$^{+1.16}_{-0.68}$
&67.5/68 &40/2 &1.25$^{+0.07}_{-0.07}$$\times 10^{40}$ &6.44$^{+7.35}_{-3.95}$$\times 10^{40}$ &78.14$^{+21.86}_{-61.43}$ $\%$  \\

...&...&...&0.15$^{+0.12}_{-0.09}$ &0.72$^{+0.05}_{-0.06}$&3.51$^{+0.92}_{-0.62}$&71.4/75 &68.4/2 &1.33$^{+0.09}_{-0.08}$$\times 10^{40}$ &3.08$^{+2.26}_{-1.22}$$\times 10^{40}$ &60.77$^{+39.23}_{-37.97}$ $\%$  \\

all$^{g}$&...&...&0.23$^{+0.06}_{-0.05}$ &0.76$^{+0.03}_{-0.04}$&3.09$^{+0.34}_{-0.31}$
&212.4/147$^{\dagger}$ &149.8/2 & | &| &|  \\

NGC 6946& 0.0001&21.1&0.45$^{+0.17}_{-0.12}$&0.28$^{+0.07}_{-0.05}$&2.26$^{+0.10}_{-0.09}$&294.5/258& 34.4/2&1.50$^{+0.05}_{-0.05}$$\times 10^{39}$ &4.26$^{+1.89}_{-1.17}$$\times 10^{39}$ &72.31$^{+6.66}_{-7.29}$ $\%$  \\

...&...&...&0.64$^{+0.25}_{-0.49}$ &0.24$^{+0.14}_{-0.05}$&2.22$^{+0.16}_{-0.17}$&75.8/79&12.2/2 & 1.44$^{+0.09}_{-0.10}$$\times 10^{39}$ &6.75$^{+6.53}_{-6.13}$$\times 10^{39}$ &46.68$^{+7.24}_{-8.06}$ $\%$  \\

\hline
\end{longtable}
{\small
\begin{flushleft}
{$^{a} \rm Total~galactic~H I~column~density~in~units~of~10^{20}~cm^{-2};  $
\\$^{b} \rm Obtained~column~density~in~units~of~10^{22}~cm^{-2}~to~estimate~redshifted~absorption~due~to~the~ISM~within~the~host~galaxy;   $
\\$^{c} \rm
Improvement~in~\chisq~of~the~absorbed~power~law~model~with~collisionally~ionized~plasma
~ over~the~absorbed~power~law~model~\newline and~reduction~ of~degrees~of~freedom~in~the~fit.$ 
\\$^{d} \rm Luminosity~in~erg~s^{-1},~(0.3-10~keV)$.
\\$^{e} \rm Few~counts~for~the~central~source$.
\\$^{f} \rm Model~with~absorption~and~two~power-law~components~$. 
\\$^{g} \rm Fit~performed~by~combining~the~datasets~available~for~the~nucleus~to~check~for~variability$.
\\$^{\star} \rm Null~hypothesis~rejected~(the~probability~of~the~\chisq~distribution~is~<<10^{-4})$.
\\$^{\dagger} \rm Null~hypothesis~questioned~(the~probability~of~the~
\chisq~distribution~is~close~to~1~or\sim 10^{-3})$.
} 
\end{flushleft}
}
\end{landscape}

\subsubsection{Physical parameters from the model fits}

\begin{figure*}[!t]
 \centering
\includegraphics[width=0.36\textwidth]{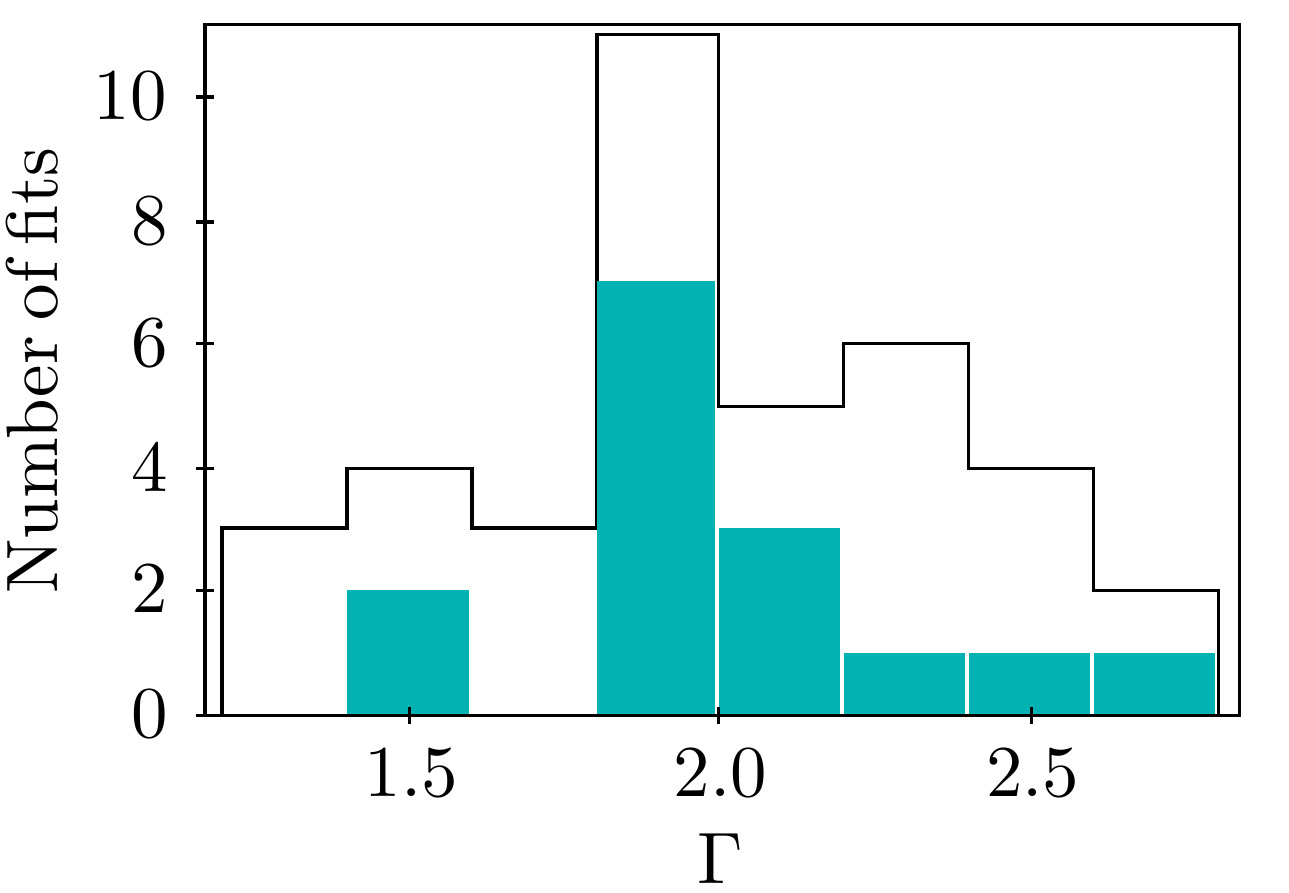}
\includegraphics[width=0.36\textwidth]{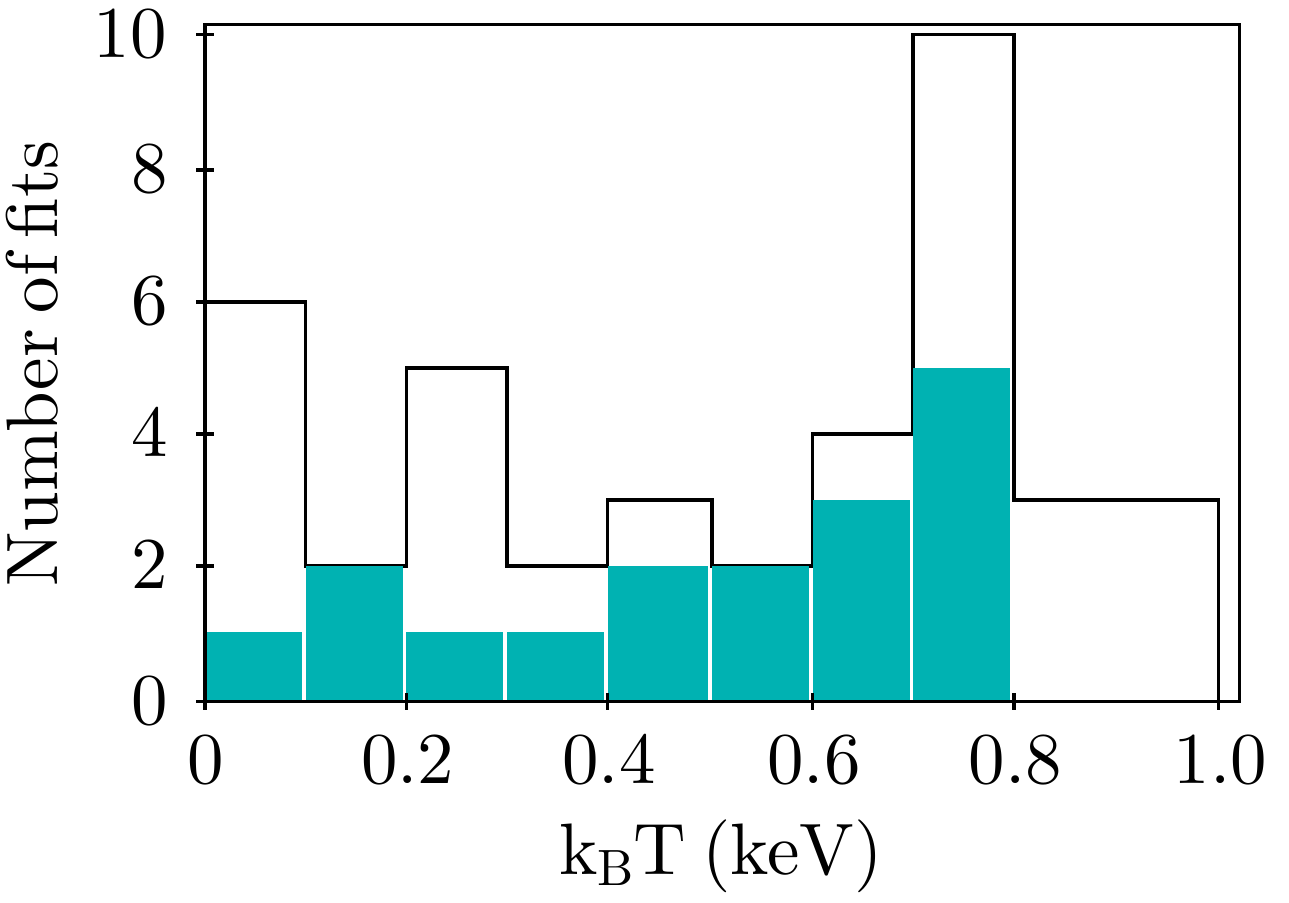}
 \includegraphics[width=0.362\textwidth]{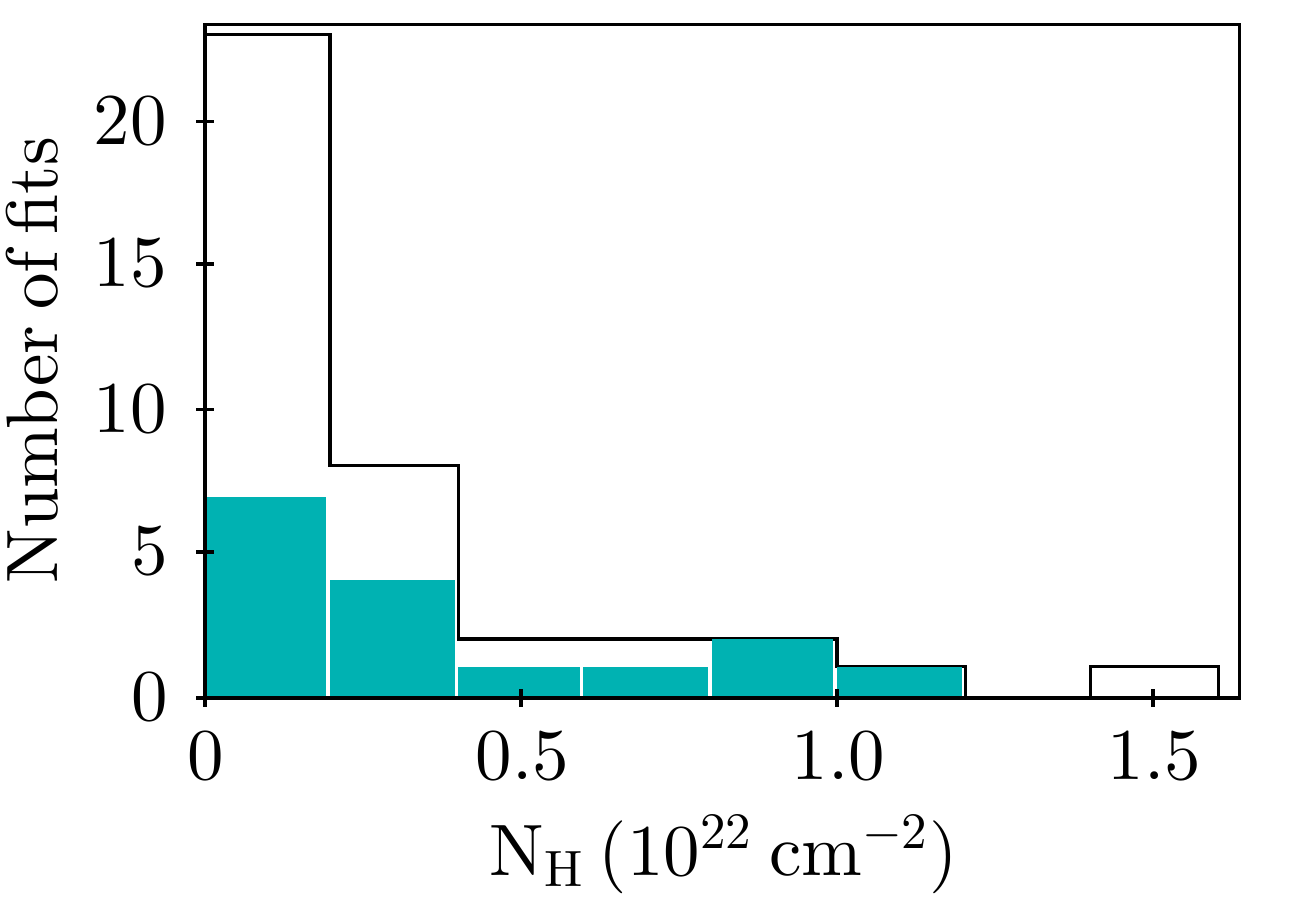}
\includegraphics[width=0.362\textwidth]{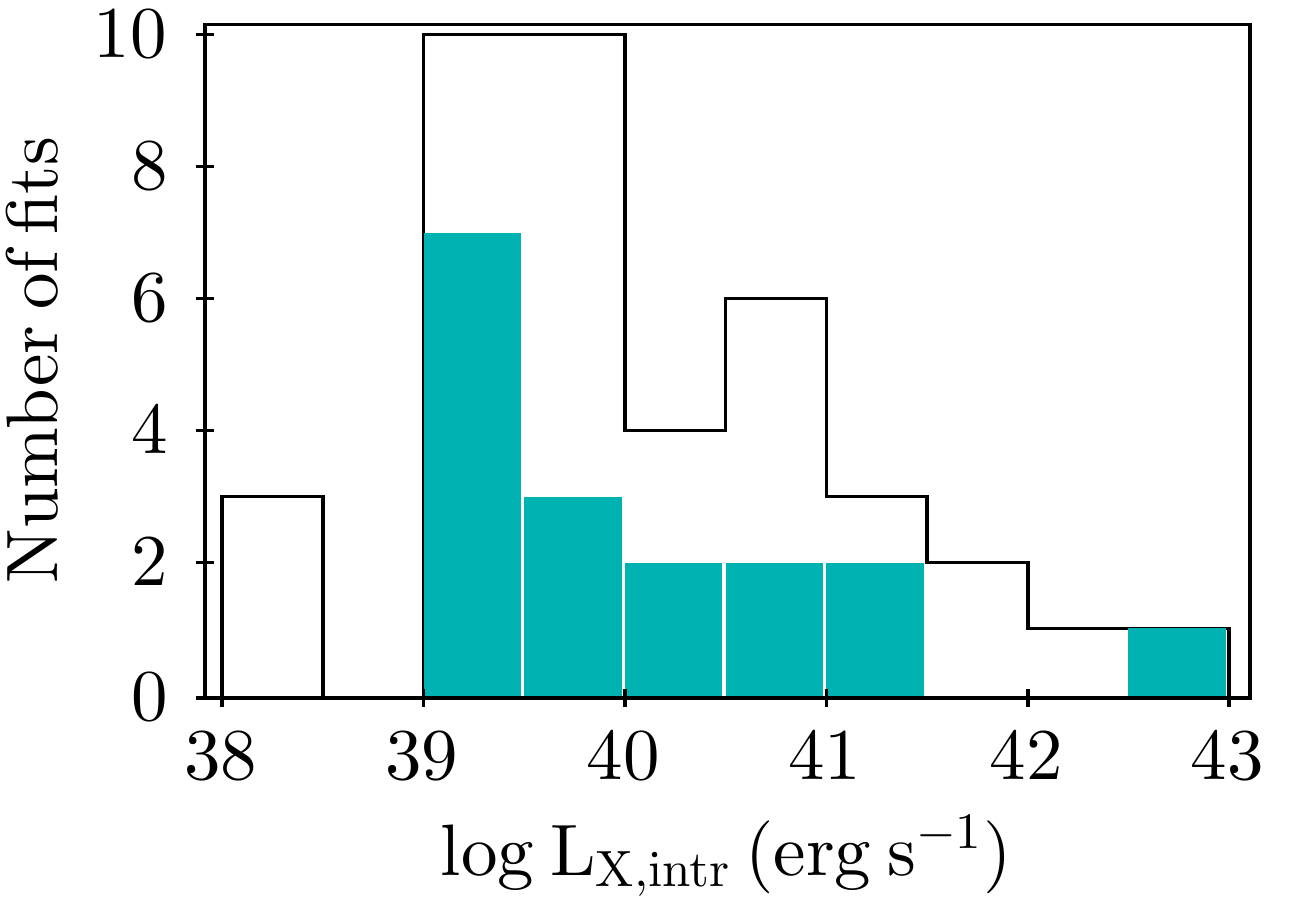}
\caption{Histogram showing the spread of obtained parameters from spectral fitting. 
The whole sample is shown (solid black line), as well as values obtained just for our LLAGN candidates (marked in blue). 
In the histogram for $N_{H}$, the heavily absorbed NGC 4845 is excluded for clarity. 
}
\label{fig:fits}
\end{figure*}

The histograms (Fig. \ref{fig:fits}) show the spread of the obtained from the model fit parameters - $N_{H,int}$, $k_{B}T$, $\Gamma$ and $L_{intrinsic}$. 
The histograms for the  LLAGN-candidates are marked in blue. 
The obtained column densities intrinsic to the host nuclei are $< 10^{22}$ cm$^{-2}$, except for NGC 4517 ($ \approx 1.5 \times 10^{22}$ cm$^{-2}$) and NGC 4845 ($ \approx 10^{23}$ cm$^{-2}$). 
The sample mean column density is log $N_{H}$ $\approx$ 21.18, with a standard deviation of 0.53.
The mean unabsorbed luminosity of our sample fits is log $L_{X} \approx$ 40.11, with a standard deviation of  1.03. 
The  low values for the column densities suggest that no type 2 LLAGN are likely to be present in our sample (except for  NGC 4845). The low obtained column densities and intrinsic luminosities, as well as a lack of Fe K$\alpha$ line detection in our sample,  would suggest that our LLAGN candidates possibly lack tori and are therefore more consistent with the ADAF/RIAF model  than with the Unified Model of AGN.

\section{Discussion}\label{discussion}

\subsection{The cross-matching}
\subsubsection{The matching statistics}
X-ray selection proved to be an efficient way of identifying accreting sources  within the Ho, Filippenko and Sargent optical sample \cite{ho:1997a, ho:1997b, ho:1997c}, as anticipated (see section \ref{reasons}). Creating a cross-match between the most up-to-date \xmm catalogue and the optical sample allowed us to survey the cores of a representative sample of nearby galaxies, due to the good coverage ($\approx$ 38\% of the galaxies are in the \xmm field of view, see Table  \ref{table:coverage}).
The well-quantified optical centre coordinates for the majority of our sample of nearby galaxies  (95\% confidence range of $\sim$ 0.5-1.5" in NED), as well as the high positional accuracy of the sources in \3xmm provided sufficiently reliable cross-matching for the purposes of this project. 
As galaxies potentially harbouring \llagns were expected to have weak or obscured X-ray fluxes, focusing on nearby galaxies ($z \lesssim$ 0.02) contributed to the selection of a sample with good X-ray photon statistics.  
Apart from the high fraction of detected Seyferts and LINERs within the \xmm field of view (49 out of 50 observed Seyferts and all 40 LINERs), there is also a  large fraction of X-ray detected  transition-LINER and \hii galaxies. 
 
The wide bandpass of the EPIC cameras (Table 1), particularly at harder energies, played a vital role in obtaining such large detected fractions. 
Whereas the high detection fraction for nuclei with AGN-dominated optical emission was anticipated, the large X-ray detection fraction for the remaining subsets suggests that some of these nuclei could possess some form of hidden in the optical regime accretion.

\subsubsection{How reliable is the cross-matching?}
The validity of the cross-matching was investigated by producing a cumulative plot of the sources and the separation between their NED and \3xmm positions. \3xmm objects which are point source matches were expected - and indeed found - to peak at very small offsets from the optical nuclei.  
Confirmed AGN and AGN-candidates in the source-separation plot were a natural check whether the match is reliable. Indeed, the optically-identified Seyferts and pure LINERs show the most distinctive peak closest to the optical centre (Fig. \ref{fig:separation}). 
Galaxies from the different optical nuclear subsets were found to have different X-ray matching fractions. 

The \3xmm high positional accuracy ($<$1.5")  allows us to minimise contamination of the nuclear sample by unrelated sources simply by looking at a very confined region, while being confident that the putative LLAGN are covered by the observations. However, it is inevitable for dense source fields to have a mixture of spatially unresolved multiple nuclear sources within the PSF of \xmm (see Table \ref{epic}), especially for late-type (star-forming) galaxies (Section \ref{sfr}). 
Without examining in detail the sources of contamination for each of the galaxies, a firm conclusion about the fraction of X-ray detected LLAGN in the subsamples cannot be drawn. 
A good estimate for the contamination per unit area can be evaluated on a galaxy-to-galaxy basis, by cross-matching all the X-ray sources within the optical extent of each galaxy, which was beyond the scope of the project.

\subsection{Implications of the results derived from the optical/X-ray matched sample}

\subsubsection{Sample luminosities}
The values of the average X-ray luminosities calculated for each nuclear subset apart from NoELs are consistent with LLAGN, but also with other X-ray - luminous sources  (Table \ref{table:lum}, also see Section \ref{sfr}). Therefore, the proximity to the nucleus is the main reason to believe that the majority of the detections are likely to be powered by a LLAGN, especially above $L_{X}$ of 10$^{40}$ \ergsec.  Our test by plotting  $L_{H\alpha}$ \cite{ho:1997a} and $L_{X}$ (0.2-10 keV) of the nuclei with point-like X-ray sources in our sample (Fig. \ref{fig:lum1}) supports this conclusion, as the majority of our optical Seyferts and LINER nuclei, as well as many of the optical \hii and transition-LINERs, lie in the 100 $> L_{X}/L_{H\alpha} >$ 1 range. 

\subsubsection{The black hole masses and accretion rates}

Black hole masses, along with accretion rates, are a fundamental property of all AGN.  Calculating the black hole masses for our sample set an upper limit to the energetics of our putative LLAGN via the Eddington limit.  The obtained average black hole masses are all consistent with SMBHs residing in the centres of the galaxies within the subsets (see Section 1.2 and Table \ref{table:eddratio}). The  Eddington ratios obtained for our subsets are sub-Eddington, consistent with inefficient accretion and therefore with the RIAF/ADAF model for the central engine of LLAGN. Taking into account the fact that 
our mass accretion rate calculation assumed a Compton-thick, physically-thin disc, this estimate serves as an approximation to the true accretion rates. 

The average  $M_{BH}$ values do not  clearly suggest downsizing  of the optically defined nuclear subsets within our  matched sample, as the average SBMH mass for Seyferts  is, in fact, lower than the estimates for the LINER groups. 
Limitations on the empirical relation used to compute $M_{BH}$ are imposed by the different dispersions, black hole masses, as well as distance errors of the galaxy sample from which it is derived, inflicting some uncertainty on the slope\cite{hobh:2000}.  
In any case, the relation would hold well for small velocity dispersions (true for the majority of the galaxies in our sample), and therefore it provides a good estimate for the SMBH masses. 
Despite the large spread of $M_{BH}$ values for a given X-ray luminosity, the overall shape of the plot in Fig.  \ref{fig:bhmass} shows that the more luminous nuclei in our sample generally contain more massive SMBHs, particularly when comparing the optical \hii galaxies with the remainder of our sample. 

\subsubsection{Sample morphology and X-ray nuclei}
Our fraction of  X-ray detected early-type galaxies ($\approx$ 64\%) is consistent with the fraction of AGN found in early-types by optical  spectroscopic studies ($\approx$ 50 - 70 \%) \cite{ho:2008, zhang:2009}. However, the fraction in the later Hubble types ($\approx$ 55\%) is much higher than inferred from these studies ($\approx$ 15\%). Again, this can be explained by the presence of faint active nuclei, which would have small or obscured contributions to the optical/UV emission.
Another reason, however, could be small number statistics, or contamination from HMXBs (high-mass X-ray binaries), found predominantly in star-forming spiral galaxies \cite{zhang:2009}. 
The spatial resolution of \xmm corresponds to a linear size   of  $\sim$300 parsecs on average in the galaxies in our sample (with a minimum of $\sim$10 pc and maximum of $\sim$1000 pc). The average scale would encompass, in many of the galaxies, not only a potential LLAGN, but also an inner bulge component (if present), and/or the star-forming regions in and around the galaxy cores \cite{roberts:2001}.  

\subsection{The spectral fits and sample separation}

\subsubsection{The tests and their limitations}
The project goal addressed by the spectral analysis of \hii and transition-LINER nuclei was to determine which mechanism is more likely to be powering the X-ray emission observed by the EPIC cameras. This was addressed by  modelling the available X-ray spectra for each central source with {\sc{xspec}}.
 The observations included in Table 7 all met the Gaussian criteria for $\chisq$ fitting.  

The initial test to investigate the origin of the X-ray emission was to compare two simple spectral models  - an absorbed power-law continuum, and an absorbed power-law continuum with a thermal component. 
 Statistical differences between the two models would generally be a good test to identify the origin of the X-ray emission in the case of high quality, complex data. However, for the fits with no statistically significant improvement in $\chisq$ of the \texttt{(apec+po)}-based model over the \texttt{po}-based model, there is a clear trend that this occurred in the cases with fewer available counts in the spectra. Therefore, this was not a reliable way to distinguish between the two models in our sample. 

The absorbed \texttt{(apec+po)} model lead to good statistics  and a physically meaningful  spread of parameters for our sample. The power-law  and  thermal components in our \xmm spectra were both readily detectable in the majority of the fitted datasets. Therefore, the next test was to investigate the relative contributions of the components of the  absorbed \texttt{(apec+po)} model to the observed emission. The relative simplicity of the model (given the good data in many cases), is the main reason for the large uncertainties on the parameters in some of the fits. 
By comparing the model components, the dominant contribution to the unabsorbed model flux was found to be due to the power-law continuum, with varying level of contribution by the fitted thermal component for each source.  
Therefore, our sample is overall consistent with unresolved point-like source continua with thermal plasma emission. The hard emission is therefore possibly from X-ray binaries and/or a hidden AGN continuum. 

The scenario of having a  source-mixture spectrum is highly likely when multiple sources are found in the higher-resolution \textit{Chandra} images within the \xmm source extraction region, whereas a true LLAGN is more probable  when a single source of high intrinsic luminosity was found very close to the optical nucleus position. For the nuclei without available \textit{Chandra} data or multiple observations to suggest variability, any distinction between a spectrum originating from a mixture of sources or one resulting from an optically-undetected LLAGN is highly ambiguous. 

Taking into account the fact that nearby starburst galaxies can have power-law contribution to the emission of up to $\sim$ 82\% (for similar to our fitting techniques, Section \ref{checks}) ruled out a large fraction of the power-law dominated sources as potential LLAGN. However, the large uncertainty on the contribution in some cases due to the error on the obtained source luminosities should be noted.
It should also be noted that the sample of 82 LINERs, on which this distinction was based, carries  selection bias. This is  due to the fact that the LINER sample is derived from  an infrared-selected  sample, and this method suffers from confusion with processes unrelated to AGN activity \cite{ho:2008}. LINER-like emission can be generated in the IR as a result of shocks and starburst activity, particularly in interacting systems \cite{ho:2008}. Therefore, our adopted limit on the thermal emission may not be very accurate.

\subsubsection{The LLAGN candidates}
Considering the limitations stated above, our methods proved to be generally reliable for the sample separation into potential LLAGN and non-LLAGN. As discussed, our transition-LINER and \hii galaxies are especially prone to a high number of X-ray binaries and other X-ray sources coexisting with a hypothetical LLAGN (see Section \ref{sfr}).  Exploring the \textit{Chandra} images and available data from studies, we conclude that contamination in our nuclear sample is very low, and therefore the luminosity output should be due to an accreting nuclear SMBH in the majority of our LLAGN candidates. Classified by other studies LLAGN (NGC 3627, NGC 4845, NGC 4102, NGC 4713, NGC 6217) 
  are found into our LLAGN candidates subsample \cite{liu:2011, zhang:2009,veron:2006}. Known from X-ray studies starburst galaxies, such as NGC 1569 \cite{heike:2004}, are excluded from the group. NGC 598, a known ULX \cite{roberts:2013}, was also excluded by our method.
 However, the  comparative nature of the fitting possibly led to the omission of some potential LLAGN. NGC 4536, for example, was the focus of a recent \xmm study \cite{satyapal:2011}, and was reported as an LLAGN, after a more detailed investigation of the possible contributions to the X-ray luminosity and considering area-density contamination.  

In the case that the sources in our candidate sample are, in fact, optically-hidden LLAGN, can we say anything about their type from the X-ray data? The lack of Fe K$\alpha$ emission detection generally would suggest that the hard X-ray spectra are not dominated by reflection from  neutral material with a high column density. The good spectral resolution of the \xmm instruments at the  Fe K$\alpha$ energy  gives cause to believe this lack of detection  in the cases with sufficient counts (see Table \ref{epic}). All the nuclear sources have soft spectra, generally with an upper limit of $\Gamma \sim$ 2, except for NGC 5055, which appears harder, with $\Gamma \lesssim$ 1.7. As well as this, the upper limits of the obtained \hi column densities do not exceed 10$^{22}$ cm$^{-2}$ (except for our confirmed Seyfert 2 - NGC 4845).  
Therefore, overall the sample is consistent with type 1 nuclei. 
Given the low column densities, if our sources are LLAGN, the cause for the elusiveness of  AGN signatures in the an optical regime most likely cannot be attributed to strong obscuration. An already discussed reason for the lack of optical  signatures is dilution by starburst activity, which is highly likely in the case of the late-type spirals in the sample and particularly for interacting systems (see Section \ref{cand}).  The other main possible scenario are low mass accretion rates, which lead to a RIAF, unable to ionize a BLR (Section 1.3).

\section{Conclusion}

In this project, the X-ray nuclear  properties of a representative sample of nearby optically-bright galaxies were explored by matching the detailed classification by Ho, Sargent and Filippenko, based on optical spectral  features, to the latest available \xmm catalogue. Investigation of the matched sample properties lead to the conclusion that they are overall consistent with the presence of a large fraction of X-ray detected \llagns with inefficient accretion within our sample. Such studied properties included the offset of the detected sources from the optical nuclei, the X-ray luminosity ($\overline{L}_{X} \approx 10^{40}$ \ergsec), the black hole masses ($\overline{M}_{BH} \sim 10^{8}$ \Msun, in the range for SMBHs), the Eddington ratios ($\overline{\lambda}_{Edd} \sim -4$, severely sub-Eddington). The Eddington ratio estimates and $L_{X}/L_{H\alpha}$ plots suggested that, 
in the case of low level of contamination of our sample, a significant fraction of the  \hii and transition-LINER subsets could harbour an optically  undetected LLAGN. This was examined by modelling the X-ray spectral products available for the majority of the  \hii and transition-LINERs with a detected by \xmm nuclear point source. If these spectra are not highly contaminated, as found for some of the nuclei by available higher-resolution \textit{Chandra} images, we conclude that  $\approx$ 43\% of the \hii nuclei and 40\% of the transition-LINER nuclei in the studied sample are optically undetected LLAGN candidates. In this scenario, the low \hi column densities and model intrinsic luminosities (and the inferred from the latter $\lambda_{Edd}$), are consistent with the RIAF model for LLAGN.

To give a final answer to the question of what fraction of these galaxies are true LLAGN hosts, more observations of the nuclei are required to check for variability  on typical for accreting SMBHs temporal scales.
Cash statistics could be used on unbinned spectral data in case of low-count spectral products for some of the nuclei in future investigations in order to obtain more datasets to test for such  variability. 
The possible relation between the host intrinsic column density along the line of sight  and the viewing angle can be explored, as well as the relation between X-ray luminosity and star formation rates in the galaxies in the sample.
Future  X-ray missions  will  allow improved photometric and spectral analysis by combining high spectral and angular resolution. Such an upcoming telescope is 
\textit{eROSITA}, due to be launched in 2015, which will allow the systematic detection of obscured or dim accreting black holes in up to 3 million nearby galaxies, thus providing a rich resource for more detailed future surveys.

\vspace{1cm}

{\footnotesize
\begin{multicols}{2}    
\bibliography{IvaylaKalchevaMPhysThesis}
\end{multicols}}

\begin{landscape}
\section*{Appendix}

{\footnotesize
\begin{tabular}{|lccllccl|}
\hline
  \multicolumn{1}{|c}{Name} &
  \multicolumn{1}{c}{$\delta$ (")} &
  \multicolumn{1}{c}{$\delta$ (pc)} &
  \multicolumn{1}{c}{X-ray core$^{\star}$} &
  \multicolumn{1}{c}{Variability} &
  \multicolumn{1}{c}{log$M_{BH}$} &
  \multicolumn{1}{c}{log$\lambda_{Edd}$} &
  \multicolumn{1}{c|}{Notes} \\
\hline
\hline
  IC 342 & 0.56 & 8.1 & I$^{a}$ &  & 6.5 & -4.3 & very slight extended emission, \hii$^{b}$/Starburst/\hii (NED)\\
  NGC 1569 & 4.81 & 37.3 & III$^{a}$ &  & 5.6 & -4.1 & Starburst Sy1$^{b}$\\
  NGC 2146 & 2.41 & 201.0 & III$^{b}$, 2  &  & 7.3 & -3.4 & position misalignment - optical nucleus extent$^{b}$/ \hii; LIRG (NED)\\
  NGC 2342 & 3.49 & 1174.6 &  &  & 7.6 & -3.1 & \hii; LIRG (NED) \\
  NGC 2903 & 2.2 & 67.1 &  &  & 6.8 & -4.4 & \hii \\
  NGC 3367 & 0.6 & 127.2 &  &  & 6.2 & -2.3 & LINER Sy (NED) \\
  NGC 3627 & 2.09 & 66.8 & I$^{a}$, III$^{b}$, 1 &  & 7.3 & -4.8 & LINER Sy2$^{b}$\\
  NGC 410 & 1.41 & 484.3 &    IV &  & 8.7 & -3.9 & soft extended emission; LINER \hii (NED)\\
  NGC 4102 & 0.37 & 30.8 & I$^{b}$, 1 &  & 7.9 & -4.1 & \hii LINER$^{b}$/\hii LINER (NED)\\
  NGC 4298 & 1.35 & 110.2 &  &  & 5.5 & -3.3 &  \hii (NED)\\
  NGC 4459 & 2.3 & 187.2 & II$^{b}$, 1 &  & 7.8 & -5.0 & \hii LINER,  slight extended emission$^{b}$/ soft band \\
  NGC 4470 & 4.18 & 636.2 &  &  & 6.8 & -4.0 & \hii$^{b}$\\
  NGC 4490 & 0.89 & 33.8 & IV$^{a}$ /I & XMM spectra & 5.7 & -1.2 & point source hard band, \hii (NED)\\
  NGC 4517 & 2.66 & 126.4 &  &  & 5.6 & -0.2 & \hii (NED) \\
  NGC 4536 & 1.12 & 72.0 &  &  & 6.7 & -3.8 & \hii Starburst\\
  NGC 4569 & 1.04 & 84.8 & I$^{b}$, 1 /I &  & 7.5 & -4.4 & LINER Sy$^{b}$\\
  NGC 4647 & 4.26 & 346.7 &  &  & 4.1 & -1.9 & \hii (NED) \\
  NGC 4654 & 0.82 & 66.8 & II$^{b}$, 1 /IV &  & 5.7 & -3.3 & offset in excess of X-ray positional error$^{b}$, \hii (NED)\\
  NGC 4713 & 1.17 & 101.3 & I$^{a}$, 1 /I &  & 4.6 & -2.2 & LINER$^{b}$/ \hii LINER (NED)\\
  NGC 4845 & 0.59 & 44.7 &  & Var. flag XMM & 7.4 & -1.5 & Sy 2 (NED)\\
  NGC 5055 & 0.69 & 24.1 & II$^{a}$,I$^{b}$, 1 /II, 2 &  & 7.2 & -4.7 & \hii LINER$^{b}$\\
  NGC 520 & 3.76 & 506.6 & II$^{b}$, 1 /II, 1 &  & 5.5 & -2.2 & pt. and ext. emission hard band/  Starburst (NED)\\
  NGC 5248 & 1.52 & 167.1 &  &  & 6.9 & -3.9 & Sy2 \hii (NED) \\
  NGC 5354 & 2.6 & 413.0 & IV &  & 8.2 & -5.2 & LINER$^{b}$ /no clear source in \textit{Chandra} image/ LINER (NED)\\
  NGC 5457 & 0.58 & 15.3 & I$^{a,b}$, 2 / I, 2 &  & 4.6 & -3.1 & \hii (NED) \\
  NGC 5746 & 1.17 & 166.4 & I$^{a}$, 1 /I, 2&  & 8.1 & -4.0 & \\ 
  NGC 5846 & 1.61 & 222.8 &  &  & 8.4 & -4.0 &  LINER \hii (NED)\\
  NGC 598 & 2.97 & 10.1 & II$^{a}$, I$^{b}$, 1 /I, 1 &  & 4.4 & -1.8 & \hii$^{b}$ (with ULX?)\\
  NGC 6217 & 0.38 & 43.5 &  & XMM spectra & 6.4 & -2.8 &\hii Sy2 (NED) \\
  NGC 660 & 3.68 & 210.4 & III$^{a,b}$, 1 &  & 7.4 & -4.4 & \hii LINER$^{b}$\\
  NGC 6946 & 3.54 & 94.3 & II$^{a}$, I$^{b}$, 3 /II, 2 &  & 6.0 & -3.3 & \hii $^{b}$/ Sy2 \hii\\
\hline
\end{tabular}
}

{\footnotesize
\begin{flushleft}
{$^{a} \rm classification~from~Zhang~et~al.~(2009);  ~^{b} \rm classification~from~Chandra~ACIS~survey~in~383~nearby~galaxies~-~Liu~(2011); ~ I~-~ point,~ II~-~point~ with ~extended,$
\\$  \rm III~-~ extended,~ IV~ - ~no~ detected~ emission~ at~ nuclear ~position.~~^{\star}\rm Number~of~Chandra~sources~within~6"~ of ~nuclear~ position.$}
\end{flushleft}
}
\end{landscape}

\end{document}